\documentclass[]{article}
\usepackage{bm}
\usepackage{amsmath}
\usepackage{amssymb}
\usepackage{cite}

\usepackage{pst-all}
\input epsf
\usepackage{epic}
\usepackage{graphicx}
\usepackage{pstricks}

\newtheorem{lemma}{Lemma}
\newtheorem{proposition}{Proposition}

\def\be{\begin{equation}}
\def\ee{\end{equation}}

\def\bl{\bm l}
\def\bx{\bm x}
\def\by{\bm y}
\def\bz{\bm z}
\def\bn{\bm n}
\def\bN{\bm N}
\def\bP{\bm P}
\def\bX{\bm X}
\def\bbt{\bm \beta}

\def\sA{\scriptscriptstyle A}
\def\sB{\scriptscriptstyle B}
\def\sAB{\scriptscriptstyle AB}

\begin{document}

\title{New tools for determining the light travel time \index{light travel time} in static, spherically symmetric spacetimes \index{spherically symmetric spacetime} beyond the order $G^2$}

\author{Pierre Teyssandier\thanks{E-mail: Pierre.Teyssandier@obspm.fr} \\
\small D\'ept SYRTE, CNRS/UMR 8630, UPMC, Observatoire de Paris,\\
\small 61 avenue de l'Observatoire, F-75014 Paris, France}

\date{}

\maketitle

\begin{abstract}

This paper is mainly devoted to the determination of the travel time of a photon as a function of the positions of the emitter and the receiver in a large class of static, spherically symmetric spacetimes. Such a function--often called time transfer function--is of crucial interest for testing metric theories of gravity in the solar system. Until very recently, this function was known only up to the second order in the Newtonian gravitational constant $G$ for a 3-parameter family of static, spherically symmetric metrics generalizing the Schwarzschild metric. We present here two procedures enabling to determine--at least in principle--the time transfer function \index{time transfer function} at any order of approximation when the components of the metric are expressible in power series of $m/r$, with $m$ being half the Schwarzschild radius of the central body and $r$ a radial coordinate. The first procedure is a direct application of an iterative method proposed several years ago for solving the Hamilton-Jacobi (or eikonal) equation satisfied by the time transfer function. The second procedure involves the iterative solution of an integro-differential equation derived from the \index{null geodesic equations} null geodesic equations. These procedures exclusively work for light rays which may be described as perturbations in powers series in $G$ of a Minkowskian null geodesic passing through the positions of the emitter and the receiver. It is shown that the two methodologies lead to the same expression for the time transfer function up to the order of $G^3$. The second procedure presents the advantage of exclusively needing elementary integrations which may be performed with any symbolic computer program whatever the order of approximation. The vector functions characterizing the direction of light propagation \index{direction of light propagation} at the points of emission and reception are derived up to the third order in $G$. The relevance of the third order terms in the time transfer function is briefly discussed for some solar system experiments.

$ $ 

\noindent PACS numbers: 04.20.-q, 04.25.-g, 04.80.Cc, 95.10.Jk

\end{abstract}

\section{Introduction} \label{intro}

Many experiments designed to test relativistic gravity involve photons travelling between an emitter and a receiver both located at a finite distance. Some of these experiments are based on the measurement of a time delay or a comparison of distant clocks, while the other ones aim to measure the gravitational \index{deflection of light} deflection of light. In spite of their differences, however, all these tests can be modelled by a single mathematical tool, namely the expression of the light travel time \index{light travel time} as a function of the positions of the emitter and the receiver for a given time of reception (or emission). Indeed, it has been shown that knowing such an expression, that we call a \index{time transfer function} time transfer function, makes possible to determine not only the frequency shift \index{frequency shift} and the Doppler-tracking \index{Doppler-tracking} between the emitter and the observer, but also the direction of light propagation \index{direction of light propagation} \cite{linet,leponcin1,teyssandier3,leponcin2,teyssandier2,bertone1,hees1}. 

The aim of the present paper is to give an overview of the two procedures which are currently at our disposal for calculating the time transfer function in static, spherically symmetric spacetimes \index{spherically symmetric spacetime} at least up to the order $G^3$, with $G$ being the Newtonian gravitational constant. The necessity of tackling the calculation of terms of order $G^3$ and beyond may be questioned since it is generally believed that the most accurate projects for testing the metric theories of gravity in the solar system, like SAGAS \cite{wolf}, ODYSSEY \cite{christophe}, LATOR \cite{turyshev } or ASTROD \cite{braxmaier} require the knowledge of the time transfer function only up to the order $G^2$, (see, e.g., \cite{minazzoli} and references therein). This reasoning neglects the fact that some so-called `enhanced' term of order $G^3$ in the time transfer function may become comparable to the `regular' term of order $G^2$, that is the term which can be estimated as const $\cdot$ $\frac{m^2}{cr_c}$, with $m$ being half the Schwarzschild radius of the central body and $r_c$ the zeroth-order distance of closest approach of the light ray \cite{ashby}. The enhancement occurs in a close superior conjunction, i.e. in the case where the emitter and the receiver are almost on the opposite sides of the central mass---a configuration of crucial importance in experimental gravitation. This effect is recovered from the full expression of the time transfer function up to order $G^3$ obtained in a recent paper \cite{linet2}. In the same work, it is shown that this term must be taken into account in solar system experiments \index{solar system experiments} aiming to determine the post-Newtonian parameter $\gamma$ with an accuracy of $10^{-8}$. Consequently, performing the calculations beyond the second order is fully relevant. We confine our exploration of the higher orders to the static, spherically symmetric spacetimes. In the present state of the art, indeed, it appears justified to neglect the relativistic contributions due to the non-sphericity or to the dynamic interactions between the Sun and the planets beyond the linear regime (see, e.g., \cite{klioner,kopeikin1,kopeikin2,linet,klioner2,kopeikin3,crosta} and refs. therein). 

We focus our attention on the theories of gravity in which it is possible to suppose that the components of the metric are analytic expansions in powers of $m/r$. The cosmological constant is neglected. The metric is thus regarded as a generalization of the Schwarzschild metric \index{Schwarzschild metric} characterized by an infinity of dimensionless constants including the well-known post-Newtonian parameters \index{post-Newtonian parameters} $\beta$ and $\gamma$. We restrict our attention to the case where the paths followed by light are what we called quasi-Minkowskian light rays \index{light ray} in \cite{teyssandier2}, namely null geodesics \index{null geodesic} described as perturbations in powers of $G$ of a Minkowskian null segment passing through the spatial positions of the emitter and of the receiver. The corresponding time transfer function is then represented by a series in powers of $G$. For the sake of brevity, a term of order $G^n$ is said to be of order $n$.

Upon these assumptions, the first-order term in the time transfer function reduces to the well-known Shapiro time delay \cite{shapiro}, which can be obtained by different reasonings, some of them involving only elementary calculations (see, e.g., \cite{will} or \cite{blanchet}). Until recently, the contributions beyond the linear regime were calculated only up to the second order. Two kinds of methods were available.

{\it a)} Integration of the \index{null geodesic equations} null geodesic equations. After the pioneering work\cite{richter}, the post-post-Newtonian expression of the time transfer function in the Schwarzschild metric \index{Schwarzschild metric} has been obtained by Brumberg \index{Brumberg V A} for a class of quasi-Galilean coordinate systems of interest in celestial mechanics \cite{brumberg1,brumberg2}. The analytic integration of the null geodesic equations in a three-parameter family of static, spherically symmetric spacetimes \index{spherically symmetric spacetime} has been recently performed in\cite{klioner1} for discussing the astrometric Gaia mission. These approaches work well, but present the drawback to be indirect, since the expression of the time transfer function is deduced from a solution which corresponds to a light ray emitted at infinity in a given direction. 

{\it b)} Methods natively adapted to the generic case where both the emitter and the receiver of the light rays are located at a finite distance from the origin of the spatial coordinates. These methods are based either on an iterative determination of the Synge world function \index{world function} (see \cite{john} for the Schwarzschild metric and \cite{leponcin1} for a three-parameter family of static, spherically symmetric spacetimes), or on an iterative integration of the Hamilton-Jacobi (or eikonal) equation satisfied by the time transfer function (see \cite{teyssandier1}, and \cite{ashby} for a more recent analysis). The two variants have been successfully employed. The results obtained by the different above-mentioned procedures are equivalent, up to a coordinate transformation.

However, the matter is currently making substantial progress with the new procedure developed in \cite{linet2}. This procedure allows to determine the time transfer function by an iterative solution of an integrodifferential equation derived from the \index{null geodesic equations} null geodesic equations. The calculations only involve elementary integrations which can be performed with any symbolic computer program whatever the order of approximation. It must be emphasized that the expression of the time transfer function up to the third order obtained in \cite{linet2} is not to be confused with the formulae found in \cite{sarmiento} and \cite{keeton} (these formulae involve the radial coordinate of the pericenter of the ray without calculating this quantity as a function of the positions of the emitter and the receiver). Moreover, this expression markedly improves the result previously given in \cite{ashby}. In \cite{ashby}, indeed, only the asymptotic form of the time transfer function when the emitter and the receiver tend to be in conjunction is found. 

Faced with such a success, it is legitimate to ask whether the iterative procedure elaborated in \cite{teyssandier1} allows to determine the time transfer function up to the third order. We prove here that this is effectively the case and that the result found in \cite{linet2} is recovered. So we have now two procedures at our disposal for a large class of static, spherically symmetric spacetimes. 
 
The paper is organized as follows. Section \ref{Notations} lists the notations and conventions we use. In section \ref{Generalities} the fundamental relations that link the light propagation direction and the frequency shift \index{frequency shift} to the time transfer function are reminded for a general static, spherically symmetric metric. In section \ref{GenAssump} the specific assumptions made on the metric and on the light rays are stated. Section \ref{FundProp} yields a recurrence relation satisfied by the perturbations terms involved in the expansion of the time transfer function. A fundamental property of analyticity is established for these terms. Section \ref{Determ1} is devoted to the first procedure presented in this paper. It is shown how the recurrence relation established in section \ref{FundProp} enables to determine an explicit expression of the time transfer up to the third order. Section \ref{Determ2} gives an overview of the procedure recently proposed in \cite{linet2}. Section \ref{pract_impl} reminds how this procedure can be noticeably simplified and leads to streamlined calculations for the time transfer function at any order. In section \ref{imp_par3} the vector functions giving the light propagation direction of a quasi-Minkowskian light ray are determined up to the third order. The results of section \ref{imp_par3} are applied in section \ref{rinfin} to a ray emitted at infinity in an arbitrary direction and observed at a given point. In section \ref{Suit3rd} the appearance of enhanced terms is rigorously proved up to the third order. The relevance of these terms for some solar system experiments is discussed in section \ref{Apexp}. Concluding remarks are given in section \ref{concl}. 

\section{Notations and conventions} \label{Notations}

We use notations and conventions as follow.

\begin{itemize}

\item
The signature of the metric is $(+,-,-,-)$.

\item
Greek indices run from 0 to 3, and latin indices run from 1 to 3.

\item
Any bold italic letter refers to an ordered triple: $(a^1,a^2,a^3)=(a^i)=\bm a$ and $(b_1,b_2,b_3)=(b_i)=\underline{\bm b}$. All the triples are regarded as 3-vectors of the ordinary Euclidean space.

\item
Given triples $\bm a$, $\bm b$, $\underline{\bm c}$, $\underline{\bm d}$, we put $\bm a . \bm b=a^i b^i$, $\bm a . \underline{\bm c}=a^i c_i$ and $\underline{\bm c} . \underline{\bm d}=c_i d_i $, with Einstein's convention on repeated indices being used.

\item
$\vert \bm a \vert$ denotes the formal Euclidean norm of the triple $\bm a$: $\vert \bm a \vert=(\bm a . \bm a)^{1/2}$. If $\vert \bm a \vert=1$, $\bm a$ is conventionally called a unit (Euclidean) 3-vector.

\item
$\bm a\times\bm b$ is the triple obtained by the usual rule giving the exterior product of two vectors of the Euclidean space.

\item
Given a bi-scalar function $F(\bx, \by)$, $\bm \nabla_{\bx}F(\bx, \by)$ and  $\bm \nabla_{\by}F(\bx, \by)$ denote the gradients of $F$ with respect to $\bx$ and $\by$, respectively.

\end{itemize}

\section{Generalities} \label{Generalities}

Before entering into the main subject of this paper, it may be useful to remind the most relevant results obtained in \cite{leponcin1} and \cite{teyssandier1} concerning the relations between the light travel time \index{light travel time} and the quantities involved in the time/frequency transfers \index{frequency transfer} experiments or in astrometry \index{astrometry}. 

Throughout this work, spacetime is assumed to be a 4-dimensional manifold endowed with a static, spherically symmetric \index{spherically symmetric metric} metric $g$. We suppose that there exists a domain ${\cal D}_h$ in which the metric is regular, asymptotically flat and may be interpreted as the gravitational field of a central body having a mass $M$. We put $m=GM/c^2$. The domain of regularity ${\cal D}_h$ is assumed to be covered by a single quasi-Cartesian coordinate system $x^{\mu}=(x^0,x^i)$ adapted to the symmetries of the metric. We use the time coordinate $t$ defined by $x^0=ct$ and we put ${\bm x}=(x^i)$, $i=1, 2, 3$. For convenience, the coordinates $(x^0,{\bm x})$ are chosen so that the metric takes an isotropic form: 
\be \label{ds2}
ds^2 = {\cal A}(r)(dx^0)^2-\frac{1}{{\cal B}(r)}\delta_{ij}dx^idx^j,
\ee
where $r=\vert\bx\vert$. Using the corresponding spherical coordinates $(r,\vartheta ,\varphi )$, one has
$$
\delta_{ij}dx^idx^j = dr^2+r^2d\vartheta^2+r^2\sin^2\vartheta d\varphi^2.
$$

We generically consider a photon emitted at a point-event $x_{\sA}$ and received at a point-event $x_{\sB}$, with $x_{\sA}$ and $x_{\sB}$ being located in the domain of regularity ${\cal D}_h$. We put $x_{\sA}=(ct_{\sA}, \bx_{\sA})$ and $x_{\sB}=(ct_{\sB}, \bx_{\sB})$. It is assumed that the photon propagates along a null geodesic \index{null geodesic} path of the metric $g$. This geodesic is denoted by  $\Gamma(\bx_{\sA},\bx_{\sB}$)\footnote{In a static spacetime, the mention of the initial time $t_{\sA}$ may be omitted.}, or simply $\Gamma$ in the absence of ambiguity. We suppose that $x_{\sA}$ and $x_{\sB}$ cannot be linked by two distinct null geodesic paths (configurations like the Einstein ring are not taken into account). Then the light travel time $t_{\sB}-t_{\sA}$ can be considered as a function of $\bx_{\sA}$ and $\bx_{\sB}$, so that one can write
\be \label{TTFa}
t_{\sB}-t_{\sA}={\cal T}(\bx_{\sA}, \bx_{\sB}; \Gamma).
\ee
We call ${\cal T}(\bx_{\sA}, \bx_{\sB}; \Gamma)$ the time transfer function \index{time transfer function} associated with $\Gamma$.

The importance of the notion of time transfer function for the astrometry \index{astrometry} and the frequency transfers \index{frequency transfer} rests on the fact that the light direction at $x_{\sA}$ and $x_{\sB}$ can be fully determined when  
${\cal T}(\bx_{\sA}, \bx_{\sB}; \Gamma)$ is explicitly known. The argument may be summarized as follows. Since the light rays \index{light ray} are null geodesic paths, the propagation direction of a photon travelling along $\Gamma(x_{\sA},x_{\sB})$ is completely characterized by the light direction triple \index{light direction triple} defined as 
\be \label{hlx}
\widehat{\underline{\bl}}_{\, x}=\left(\frac{l_i}{l_0}\right)_{x},
\ee
where $x$ denotes a point of $\Gamma(x_{\sA},x_{\sB})$ and the quantities $l_{\alpha}$ are the covariant components of a 4-vector tangent to $\Gamma(x_{\sA},x_{\sB})$ at $x$. The value of $\widehat{\underline{\bl}}_{\, x}$ is independent of the parameter describing $\Gamma(x_{\sA},x_{\sB})$. Denote by $\widehat{\underline{\bl}}_{\sA}$ and $\widehat{\underline{\bl}}_{\sB}$ the values of $\widehat{\underline{\bl}}_{x}$ at points $x_{\sA}$ and $x_{\sB}$, respectively. It is shown in \cite{leponcin1} that $\widehat{\underline{\bl}}_{\sA}$ and $\widehat{\underline{\bl}}_{\sB}$ can be inferred from the time transfer function ${\cal T}$ by using the relations 
\begin{eqnarray}
&&\widehat{\underline{\bl}}_{\,\sA}=\widehat{\underline{\bl}}_{\, e}(\bx_{\sA}, \bx_{\sB};\Gamma), \label{hlA} \\
&&\widehat{\underline{\bl}}_{\,\sB}=\widehat{\underline{\bl}}_{\, r}(\bx_{\sA}, \bx_{\sB};\Gamma), \label{hlB}
\end{eqnarray}
where the functions $\widehat{\underline{\bl}}_{\, e}$ and $\widehat{\underline{\bl}}_{\, r}$ are defined as
\begin{eqnarray} 
&&\widehat{\underline{\bl}}_{\, e}(\bx_{\sA}, \bx_{\sB};\Gamma)=c\bm \nabla_{\bx_{\sA}}{\cal T}(\bx_{\sA}, \bx_{\sB}; \Gamma), \label{hle} \\
&&\widehat{\underline{\bl}}_{\, r}(\bx_{\sA}, \bx_{\sB};\Gamma)=-c\bm \nabla_{\bx_{\sB}}{\cal T}(\bx_{\sA}, \bx_{\sB}; \Gamma). \label{hlr}
\end{eqnarray}

We can conclude from (\ref{hle}) and (\ref{hlr}) that knowing the time transfer function \index{time transfer function} associated to a given null geodesic is extremely useful in \index{astrometry} astrometry. Let us briefly examine the problem of modelling frequency shifts. Let $u_{\sA}^{\alpha}$ and $u_{\sB}^{\alpha}$ be the unit 4-velocity vectors of the emitter at $x_{\sA}$ and of the receiver at $x_{\sB}$, respectively. Denote by $\nu_{\sA}$ the frequency of the signal emitted at $x_{\sA}$ as measured by a standard clock comoving with the emitter and by  $\nu_{\sB}$ the frequency of the signal received at $x_{\sB}$ as measured by a standard clock comoving with the receiver. The ratio $\nu_{\sB}/\nu_{\sA}$ is given by the well-known formula \cite{synge}
\be \label{nuABa}
\frac{\nu_{\sB}}{\nu_{\sA}}=\frac{u_{\sB}^{\beta}(l_{\beta})_{\sB}}{u_{\sA}^{\alpha}(l_{\alpha})_{\sA}}.
\ee
Denote by $\bbt_{\sA}$ the coordinate velocity  divided by $c$ of the emitter at the instant of emission and by $\bbt_{\sB}$ the coordinate velocity  divided by $c$ of the receiver at the instant of reception, namely the triples defined as
\[
\bbt_{\sA}=\left(\frac{d\bx_{\sA}(t)}{cdt}\right)_{x_{\sA}}, \qquad \bbt_{\sB}=\left(\frac{d\bx_{\sB}(t)}{cdt}\right)_{x_{\sB}}.
\]
Noting that $l_0$ is conserved along a geodesic of (\ref{ds2}), it is immediately seen that (\ref{nuABa}) may be written in the form
\be \label{nuABl}
\frac{\nu_{\sB}}{\nu_{\sA}}=\frac{\sqrt{{\cal A}(r_{\sA})-{\cal B}^{-1}(r_{\sA})\bbt_{\sA}^2}}{\sqrt{{\cal A}(r_{\sB})-{\cal B}^{-1}(r_{\sB})\bbt_{\sB}^2}}\frac{1+\bbt_{\sB}.\widehat{\underline{\bl}}_{\, r}(\bx_{\sA}, \bx_{\sB};\Gamma)}{1+\bbt_{\sA}.\widehat{\underline{\bl}}_{\, e}(\bx_{\sA}, \bx_{\sB};\Gamma)},
\ee
where $\widehat{\underline{\bl}}_{\, e}$ and $\widehat{\underline{\bl}}_{\, r}$ are given by the right-hand side of  (\ref{hle}) and (\ref{hlr}), respectively. Formula (\ref{nuABl}) completes the proof of the relevance of the time transfer functions for experimental gravitation.

From a theoretical point of view, the problem of determining the time transfer functions in a given space-time is inextricably complicated. Indeed, given two spatial positions $\bx_{\sA}$ and $\bx_{\sB}$, and an instant $t_{\sA}$, there exists in general an infinity of light rays emitted at the point-event $(ct_{\sA},\bx_{\sA})$ and passing through point-events located at $\bx_{\sB}$. Rigorously established a long time ago for the exact Schwarzschild metric \index{Schwarzschild metric} (see, e.g., \cite{darwin}, \cite{luminet} and refs. therein), this feature occurs in a very large class of space-times \cite{giannoni}. Moreover, the full expressions of the different functions ${\cal T}(\bx_{\sA}, \bx_{\sB}; \Gamma)$ are unknown, even for the Schwarzschild spacetime. Fortunately, the gravitational field in the solar system may be regarded as weak, so that it may be assumed that the photons involved in experiments propagate along what we call quasi-Minkowskian light rays (see subsection \ref{TTFqM}). We shall see below that the corresponding time transfer function is then unique and may be determined by iterative procedures whatever the required order of approximation.


\section{Specific assumptions on the metric and the light rays} \label{GenAssump}

\subsection{Post-Minkowskian expansion of the metric} \label{TTFsss}

Metric (\ref{ds2}) is considered as a generalization of the exterior \index{Schwarzschild metric} Schwarzschild metric, which may be written in the form
\be \label{sm}
ds_{Sch}^2=\frac{\left(1-\displaystyle \frac{m}{2r} \right)^2}{\left(1+\displaystyle \frac{m}{2r}\right)^2} (dx^0)^2-
\left( 1+\frac{m}{2r}\right)^4\delta_{ij}dx^idx^j
\ee
in the region outside the event horizon located at $r=m/2$. So we henceforth assume that there exists a value $r_{h}>0$ of the radial coordinate such that the domain of regularity ${\cal D}_h$ is the region outside the sphere of radius $r_h$.  If there exists at least one event horizon, we must take for $r_{h}$ the value of $r$ on the outer horizon. By analogy with general relativity we consider that $r_h\sim m$ and we suppose that whatever $r>r_h$, ${\cal A}(r)$ and ${\cal B}^{-1}(r)$ are positive functions represented by analytical expansions as follow:
\begin{eqnarray}
&&\hspace{-10mm}{\cal A}(r) = 1-\frac{2m}{r}+2\beta \frac{m^2}{r^2}-\frac{3}{2}\beta_3\frac{m^3}{r^3}+\beta_4\frac{m^4}{r^4}+
\sum_{n=5}^{\infty} \frac{(-1)^nn}{2^{n-2}}\beta_n \frac{m^n}{r^n},\label{ppnA} \\
&&\hspace{-10mm}\frac{1}{{\cal B}(r)} = 1+2\gamma\frac{m}{r}+\frac{3}{2}\epsilon \frac{m^2}{r^2}+\frac{1}{2}\gamma_3 \frac{m^3}{r^3}+\frac{1}{16}\gamma_4 \frac{m^4}{r^4}+\sum_{n=5}^{\infty}(\gamma_n -1)\frac{m^n}{r^n}, \label{ppnB}
\end{eqnarray}
where the coefficients $\beta,\beta_3,\dots ,\beta_n,\gamma ,\epsilon ,\gamma_3,\dots ,\gamma_n,\dots$ are generalized post-Newtonian parameters \index{post-Newtonian parameters} chosen so that
\be \label{GRbc}
\beta =\gamma =\epsilon =1, \qquad \beta_n=\gamma_n=1 \quad \mbox{for} \quad n\geq3
\ee
in general relativity. 

The light rays \index{light ray} of the metric (\ref{ds2}) are also the light rays of any metric $d\tilde{s}^2$ conformal to (\ref{ds2}). This feature enables us to carry out our calculations for a metric containing only one potential. We choose $d\tilde{s}^2={\cal A}^{-1}(r)ds^2$, that is
\be \label{cds2}
d\tilde{s}^2=(dx^0)^2-{\cal U}(r)\delta_{ij}dx^idx^j,
\ee
where ${\cal U}$ is defined by
\be \label{Ur}
{\cal U}(r)=\frac{1}{{\cal A}(r){\cal B}(r)}.
\ee
It results from (\ref{ppnA}) and (\ref{ppnB}) that the potential ${\cal U}(r)$ occurring in (\ref{cds2}) may be written as 
\be \label{invAB}
{\cal U}(r)= 1+2(1+\gamma)\frac{m}{r}+\sum_{n=2}^{\infty} 2\kappa_{n}\frac{m^n}{r^n}
\ee
for $r>r_h$, with the coefficients $\kappa_{n}$ being constants which can be expressed in terms of the generalized post-Newtonian parameters \index{post-Newtonian parameters} involved in the expansions of ${\cal A}(r)$ and ${\cal B}(r)$. Taking into account a notation already introduced in \cite{teyssandier2}, namely
\be \label{kappa}
\kappa=2(1+\gamma)-\beta+\mbox{$\frac{3}{4}$}\epsilon,
\ee
$\kappa_2$ and $\kappa_3$ are given by
\be \label{kappa3}
\kappa_2= \kappa,\quad
\kappa_3=2\kappa-2\beta(1+\gamma)+\mbox{$\frac{3}{4}$}\beta_3+\mbox{$\frac{1}{4}$}\gamma_3.   
\ee

\subsection{Time transfer function for a quasi-Minkowskian light ray} \label{TTFqM}

In this paper, we restrict our attention to the special class of null geodesic \index{null geodesic} paths we have called the quasi-Minkowskian light rays \index{light ray} in \cite{teyssandier2}. This means that in what follows, the path covered by the photon is assumed to be entirely confined in ${\cal D}_{h}$ and to be described by parametric equations of the form 
\begin{eqnarray}  
&&x^0=ct_{\scriptscriptstyle A}+\xi\vert\bm x_{\scriptscriptstyle B}-\bm x_{\scriptscriptstyle A}\vert+\sum_{n=1}^{\infty}X^0_{(n)}(\bm x_{\scriptscriptstyle A}, \bm x_{\scriptscriptstyle B}, \xi), \label{qM0} \\
&&\bx=\bz(\xi)+\sum_{n=1}^{\infty}\bm X_{(n)}(\bm x_{\scriptscriptstyle A}, \bm x_{\scriptscriptstyle B}, \xi),  \label{qMi}
\end{eqnarray}
where $\xi$ is the affine parameter varying on the range $0\leq \xi \leq 1$, $\bz(\xi)$ is defined by
\be \label{zxi}
\bz(\xi)=\bx_{\scriptscriptstyle A}+\xi (\bx_{\scriptscriptstyle B}-\bx_{\scriptscriptstyle A})
\ee
and the functions $X^0_{(n)}$ and $\bm X_{(n)}$ are terms of order $n$ obeying the boundary conditions 
\begin{eqnarray} 
&&X^0_{(n)}(\bm x_{\scriptscriptstyle A}, \bm x_{\scriptscriptstyle B}, 0)=0, \label{bcd0} \\ 
&&\bm X_{(n)}(\bm x_{\scriptscriptstyle A}, \bm x_{\scriptscriptstyle B}, 0)=\bm X_{(n)}(\bm x_{\scriptscriptstyle A}, \bm x_{\scriptscriptstyle B}, 1)=0. \label{bcd}
\end{eqnarray}
According to a notation already introduced in \cite{teyssandier2}, such a null geodesic \index{null geodesic} path will be denoted by $\Gamma_{s}(\bx_{\sA},\bx_{\sB})$. For the sake of brevity, the time transfer function \index{time transfer function} associated with $\Gamma_{s}(\bx_{\sA},\bx_{\sB})$ will be henceforth denoted by ${\cal T}(\bx_{\sA},\bx_{\sB})$ or simply by ${\cal T}$. Setting $\xi=1$ in (\ref{qM0}), it may be seen that this function can be expanded in power series of $G$ as follows:
\be \label{expT}
{\cal T}(\bx_{\sA},\bx_{\sB})=\frac{\vert \bx_{\sB}-\bx_{\sA} \vert}{c}+\sum_{n=1}^{\infty}{\cal T}^{(n)}(\bx_{\sA},\bx_{\sB}) ,
\ee
where ${\cal T}^{(n)}$ stands for the term of order $n$. 

Expansion (\ref{expT}) is easy to determine when $\bx_{\sA}$ and $\bx_{\sB}$ are linked by a radial null geodesic entirely lying in ${\cal D}_h$. In this case, indeed, it is immediately deduced from (\ref{cds2}) that the expression of ${\cal T}$ is given by the exact formula
\be \label{TABrd}
{\cal T}(r_{\sA},r_{\sB}) = \mbox{sgn$(r_{\sB}-r_{\sA})$}\frac{1}{c}\int_{r_{\sA}}^{r_{\sB}} \sqrt{{\cal U}(r)}dr,
\ee
where $r_{\sA}=\vert\bx_{\sA}\vert$ and $r_{\sB}=\vert\bx_{\sB}\vert$. Substituting for ${\cal U}(r)$ from (\ref{invAB}) into (\ref{TABrd}) shows that ${\cal T}$ may be expanded as follows:  
\be \label{expTr}
{\cal T}(r_{\sA},r_{\sB})=\frac{\vert r_{\sB}-r_{\sA}\vert}{c}+\sum_{n=1}^{\infty}{\cal T}^{(n)}(r_{\sA},r_{\sB}),
\ee
where the first three perturbation terms are given by
\begin{eqnarray} 
&&\hspace{-5mm}{\cal T}^{(1)}(r_{\sA},r_{\sB}) =\frac{(1+\gamma)m}{c}\left\vert \ln\frac{r_{\sB}}{r_{\sA}}\right\vert, \label{Tr1} \\
&&\hspace{-5mm}{\cal T}^{(2)}(r_{\sA},r_{\sB})=\left[\kappa-\mbox{$\frac{1}{2}$}(1+\gamma)^2\right]\frac{m^2}{r_{\sA} r_{\sB}}\frac{\vert r_{\sB}-r_{\sA}\vert}{c}, \label{Tr2} \\
&&\hspace{-5mm}{\cal T}^{(3)}(r_{\sA},r_{\sB})=\mbox{$\frac{1}{2}$}\left[\kappa_3-(1+\gamma)\kappa+\mbox{$\frac{1}{2}$}(1+\gamma)^3\right]\frac{m^3}{r_{\sA} r_{\sB}}\left(\frac{1}{r_{\sA}}+\frac{1}{r_{\sB}}\right)\frac{\vert r_{\sB}-r_{\sA}\vert}{c}. \nonumber\\
&&\label{Tr3}
\end{eqnarray}  

Determining the right-hand side of (\ref{expT}) is much more complicated when $\Gamma_{s}(\bx_{\sA},\bx_{\sB})$ is not a radial geodesic. As it has been recalled in introduction, the perturbations terms ${\cal T}^{(n)}$ might be obtained by an iterative integration of the \index{null geodesic equations} null geodesic equations. Indeed, taking into account that $d\tilde{s}^2=0$ along a \index{null geodesic} null geodesic, it results from (\ref{cds2}) and (\ref{qMi}) that the time transfer function is given by  
\be \label{TABg}
{\cal T}(\bx_{\sA},\bx_{\sB})=\frac{1}{c}\int_{0}^{1}\sqrt{{\cal U}(r(\xi))}\left\vert  \bx_{\sB}-\bx_{\sA}+\sum_{n=1}^{\infty}\frac{d\bX_{(n)}(\bm x_{\scriptscriptstyle A}, \bm x_{\scriptscriptstyle B}, \xi)}{d\xi}\right\vert d\xi,
\ee
where the integral is taken along $\Gamma_{s}(\bx_{\sA},\bx_{\sB})$. Taking into account the boundary conditions (\ref{bcd}), it may be inferred from (\ref{TABg}) that each function ${\cal T}^{(n)}$ is theoretically calculable if the perturbations terms $\bm X_{(1)},\dots, \bm X_{(n-1)}$ involved in (\ref{qMi}) are determined by solving the null geodesic equations. This procedure is cumbersome, however. Fortunately, more workable methods can be developed, as we shall see in the next sections. 

\section{Fundamental properties of the functions ${\cal T}^{(n)}$} \label{FundProp}

\subsection{Recurrence relation satisfied by the functions ${\cal T}^{(n)}$} \label{recurTT}

Let $\bx$ be an arbitrary spatial position such that $\bx\neq\bx_{\sA}$. Consider a quasi-Minkowskian light ray $\Gamma_s(\bx_{\sA}, \bx)$ joining $\bx_{\sA}$ and $\bx$. The covariant components of a vector tangent to $\Gamma_s(\bx_{\sA}, \bx)$ at $\bx$ satisfy the equation
\be \label{iso}
(l_0)^2_x-{\cal U}^{-1}(r)\delta^{ij} (l_i)_x (l_j)_x=0 
\ee
since $\Gamma_s(\bx_{\sA}, \bx)$ is a null geodesic of metric (\ref{cds2}). Dividing (\ref{iso}) side by side by $[(l_0)_x]^2$, and then taking into account (\ref{hlr}), it is easily seen that ${\cal T}(\bx_{\sA},\bx)$ satisfies an eikonal equation as follows
\be \label{HJ}
c^2 \vert\bm \nabla_{\bx}{\cal T}(\bx_{\sA},\bx)\vert^2 = {\cal U}(r).
\ee
This equation could be solved by applying the iterative procedure developped in \cite{teyssandier1} for a general metric. Nevertheless, this procedure is so simple for an eikonal equation like (\ref{HJ}) that a specific proof deserves to be explicited as follows. 

Replacing ${\cal T}(\bx_{\sA},\bx)$ by its expansion in powers of $G$ and ${\cal U}(r)$ by (\ref{invAB}), it is immediately seen that equation (\ref{HJ}) is equivalent to the infinite system of equations
\begin{eqnarray} 
&&\hspace{-15mm}c\frac{\bx-\bx_{\sA}}{\vert\bx-\bx_{\sA}\vert} .\bm \nabla_{\bx}{\cal T}^{(1)}(\bx_{\sA},\bx)=(1+\gamma)\frac{m}{r}, \label{edpT1}\\
&&\hspace{-15mm}c\frac{\bx-\bx_{\sA}}{\vert\bx-\bx_{\sA}\vert}.\bm \nabla_{\bx}{\cal T}^{(n)}(\bx_{\sA},\bx)=\kappa_n\frac{m^n}{r^n}\nonumber \\
&&\qquad\qquad\quad-\frac{c^2}{2}\sum_{p=1}^{n-1}\bm \nabla_{\bx}{\cal T}^{(p)}(\bx_{\sA},\bx).\bm \nabla_{\bx}{\cal T}^{(n-p)}(\bx_{\sA},\bx) \label{edpTn}
\end{eqnarray}
for $n\geq2$. This system is valid for any point $\bx$. Consequently, we may suppose that $\bx=\bz(\xi)$, with $\bz(\xi)$ being defined by (\ref{zxi}), which means that $\bx$ is varying along the straight segment joining $\bx_{\sA}$ and $\bx_{\sB}$. Then we have for any $n\geq1$
\be \label{stl}
\left[\frac{\bx-\bx_{\sA}}{\vert\bx-\bx_{\sA}\vert}.\bm \nabla_{\bx}{\cal T}^{(n)}(\bx_{\sA},\bx)\right]_{\bx=\bz(\xi)}=\bN_{\!\sA\sB}.\left[\bm \nabla_{\bx}{\cal T}^{(n)}(\bx_{\sA},\bx)\right]_{\bx=\bz(\xi)},
\ee
where $\bN_{\!\sA\sB}$ is defined by 
\be \label{NAB}
\bN_{\!\sA\sB}=\frac{\bx_{\sB}-\bx_{\sA}}{\vert\bx_{\sB}-\bx_{\sA}\vert}.
\ee
But a straightforward calculation shows that  
\be \label{ttder}
\bN_{\!\sA\sB}.\left[\bm \nabla_{\bx}{\cal T}^{(n)}(\bx_{\sA},\bx)\right]_{\bx=\bz(\xi)}=\frac{1}{\vert \bx_{\sB} -\bx_{\sA} \vert}\frac{d}{d\xi}{\cal T}^{(n)}(\bx_{\sA},\bz(\xi)),
\ee 
where $d{\cal T}^{(n)}(\bx_{\sA},\bz(\xi))/d\xi$ denotes the total derivative of ${\cal T}^{(n)}(\bx_{\sA},\bz(\xi))$ with respect to $\xi$ along the segment joining $\bx_{\sA}$ and $\bx_{\sB}$. Consequently, the system of equations (\ref{edpT1})-(\ref{edpTn}) may be written in the form
\begin{eqnarray}
&&\hspace{-10mm}\frac{d}{d\xi}{\cal T}^{(1)}(\bx_{\sA},\bz(\xi))=\frac{1}{c}\vert \bx_{\sB} -\bx_{\sA} \vert \frac{(1+\gamma)m}{\vert\bz(\xi)\vert},\label{eqdT1} \\
&&\hspace{-10mm}\frac{d}{d\xi}{\cal T}^{(n)}(\bx_{\sA},\bz(\xi))=\frac{1}{c}\vert \bx_{\sB} -\bx_{\sA} \vert\Bigg\lbrace\kappa_n\frac{m^n}{\vert\bz(\xi)\vert^n}\nonumber\\
&&\qquad\quad-\frac{c^2}{2}\sum_{p=1}^{n-1}\left[\bm \nabla_{\bx}{\cal T}^{(p)}(\bx_{\sA},\bx).\bm \nabla_{\bx}{\cal T}^{(n-p)}(\bx_{\sA},\bx)\right]_{\bx=\bz(\xi)}\Bigg\rbrace
\label{eqdTn}
\end{eqnarray}
for $n\geq2$. Integrating eqs. (\ref{eqdT1}) and (\ref{eqdTn}) on the range $0\leq\xi\leq1$ and noting that ${\cal T}^{(n)}(\bx_{\sA},\bx_{\sA})=0$, we get the fundamental proposition which follows. 

\begin{proposition} \label{itTAB}
The perturbation terms ${\cal T}^{(n)}$ involved in expansion (\ref{expT}) may be written in the form
\be \label{TnFn}
{\cal T}^{(n)}(\bx_{\sA},\bx_{\sB})=\frac{1}{c}\vert \bx_{\sB} -\bx_{\sA} \vert F^{(n)}(\bx_{\sA},\bx_{\sB}),
\ee 
where the functions $F^{(n)}$ are determined by the recurrence relation 
\begin{eqnarray}  
& &\hspace{-10mm}F^{(1)}(\bx_{\sA},\bx_{\sB})=(1+\gamma)m \int_{0}^{1}\frac{d\xi}{\vert \bz(\xi)\vert}, \label{FAB1} \\
& &\hspace{-10mm}F^{(n)}(\bx_{\sA},\bx_{\sB})=\kappa_n m^n \int_{0}^{1}\frac{d\xi}{\vert \bz(\xi)\vert^n} \nonumber \\
& &\quad\quad-\frac{c^2}{2}\int_{0}^{1}\sum_{p=1}^{n-1}\left[\bm \nabla_{\bx}{\cal T}^{(p)}(\bx_{\sA},\bx).\bm \nabla_{\bx}{\cal T}^{(n-p)}(\bx_{\sA},\bx)\right]_{\bx=\bz(\xi)} d\xi \label{FABn}
\end{eqnarray}
for $n\geq2$, with the integrals being taken along the segment defined by the parametric equation 
$\bx=\bz(\xi), 0\leq\xi\leq1$.
\end{proposition} 

The recurrence relation explicited in proposition \ref{itTAB} shows that $\Gamma_{s}(\bx_{\sA},\bx_{\sB})$ is unique provided that expansion (\ref{expT}) is an admissible representation of the time transfer function. However, determining the most general conditions under which our construction is valid remains an open problem. According to (\ref{FAB1}) and (\ref{FABn}), the functions $F^{(n)}$ are given by integrals involving the analytic expansion of the metric along the straight segment joining $\bx_{\sA}$ and $\bx_{\sB}$ (see also \cite{teyssandier1}). Consequently, we shall henceforth assume that the expression $\vert \bx_{\sB}-\bx_{\sA}\vert [1+\sum_{p=1}^{n} F^{(p)}(\bx_{\sA},\bx_{\sB})]/c$ constitutes a reliable approximation of the time transfer function as long as the straight segment joining $\bx_{\sA}$ and $\bx_{\sB}$ does not intersect the hypersurface $r=r_h$, a condition expressed by the inequality
\be \label{strl}
\vert \bz(\xi) \vert >r_h \quad \mbox{for}\quad 0\leq \xi \leq 1.
\ee
We shall see in section \ref{Suit3rd} that this condition is largely satisfied by a light ray emitted in the solar system (or coming from a star) and observed in the solar system after having grazed the Sun.

\subsection{Analyticity of the functions ${\cal T}^{(n)}$} \label{analytTT}

A property of analyticity which is indispensable for justifying the procedure developed in sections \ref{Determ2} and \ref{pract_impl} can easily be inferred from proposition \ref{itTAB}. Let us begin with proving the following lemma.

\begin{lemma} \label{analytF}
The functions $F^{(n)}$ recursively determined by (\ref{FAB1}) et (\ref{FABn}) are analytic in $\bx_{\sA}$ and $\bx_{\sB}$, except when $\bx_{\sA}$ and $\bx_{\sB}$ are such that $\bn_{\sB}=-\bn_{\sA}$, with $\bn_{\sA}$ and $\bn_{\sB}$ being defined as 
\be \label{nAnB}
\bn_{\sA}=\frac{\bx_{\sA}}{r_{\sA}}, \quad \bn_{\sB}=\frac{\bx_{\sB}}{r_{\sB}}. 
\ee.  
\end{lemma}

{\bf Proof of lemma \ref{analytF}.} The proposition is obviously true for $n=1$, since the integrand $1/\vert \bz(\xi)\vert$ in (\ref{FAB1}) is analytic in $\bx_{\sA}$ and $\bx_{\sB}$ for any $\xi$ such that $0\leq\xi\leq1$ provided that $\bn_{\sB}\neq -\bn_{\sA}$. Suppose now the validity of lemma \ref{analytF} for $F^{(1)},\dots,F^{(n)}$. Assuming $p$ to be such that $1\leq p\leq n$, and then substituting $\bz(\xi)$ for $\bx$ into $\bm \nabla_{\bx}{\cal T}^{(p)}(\bx_{\sA},\bx)$, it is immediately inferred from (\ref{TnFn}) that 
\begin{eqnarray} 
&&c\left[\bm \nabla_{\bx}{\cal T}^{(p)}(\bx_{\sA},\bx)\right]_{\bx=\bz(\xi)}=\bN_{\!\sA\sB}F^{(p)}(\bx_{\sA},\bz(\xi)) \nonumber \\
&&\qquad\qquad\qquad\qquad\qquad+\xi\vert\bx_{\scriptscriptstyle B}-\bx_{\scriptscriptstyle A}\vert\left[\bm \nabla_{\bx}F^{(p)}(\bx_{\sA},\bx)\right]_{\bx=\bz(\xi)}.\label{nabpT}
\end{eqnarray}
Using (\ref{nabpT}) leads to
\begin{eqnarray}  
& &\hspace{-8mm}c^2\left[\bm \nabla_{\bx}{\cal T}^{(p)}(\bx_{\sA},\bx).\bm \nabla_{\bx}{\cal T}^{(n+1-p)}(\bx_{\sA},\bx)\right]_{\bx=\bz(\xi)}\nonumber \\
&&\quad=F^{(p)}(\bx_{\sA},\bz(\xi))F^{(n+1-p)}(\bx_{\sA},\bz(\xi))\nonumber \\
& &\quad\quad+\xi (\bx_{\sB}-\bx_{\sA}).\left[F^{(p)}(\bx_{\sA},\bx)\bm \nabla_{\bx}F^{(n+1-p)}(\bx_{\sA},\bx)\right. \nonumber \\
& &\quad\quad+\left. F^{(n+1-p)}(\bx_{\sA},\bx)\bm \nabla_{\bx}F^{(p)}(\bx_{\sA},\bx)\right]_{\bx=\bz(\xi)} \nonumber \\
& &\quad\quad+\xi^2\vert\bx_{\sB}-\bx_{\sA}\vert^2\left[\bm \nabla_{\bx}F^{(p)}(\bx_{\sA},\bx).\bm \nabla_{\bx}F^{(n+1-p)}(\bx_{\sA},\bx)\right]_{\bx=\bz(\xi)}.\hspace{3mm} \label{naTnaT}
\end{eqnarray}
It follows from our assumption that the right-hand side of (\ref{naTnaT}) is a sum of functions which are analytic in $\bx_{\sA}$ and $\bx_{\sB}$ for any $\xi$ such that $0\leq\xi\leq1$, except if $\bn_{\sB}=-\bn_{\sA}$. Each integral 
\[
\int_{0}^{1}\left[\bm \nabla_{\bx}{\cal T}^{(p)}(\bx_{\sA},\bx).\bm \nabla_{\bx}{\cal T}^{(n+1-p)}(\bx_{\sA},\bx)\right]_{\bx=\bz(\xi)}d\xi
\]
is therefore analytic if $\bn_{\sB}\neq -\bn_{\sA}$. The same property is obviously possessed by the integral $\int_{0}^{1}d\xi/\vert\bz(\xi)\vert^{n+1}$. Lemma \ref{analytF} is thus proved by recurrence.

Since $\vert\bx_{\scriptscriptstyle B}-\bx_{\scriptscriptstyle A}\vert$ is analytic except if $\bx_{\scriptscriptstyle B}\neq\bx_{\scriptscriptstyle A}$, we can state the proposition below.

\begin{proposition} \label{analytT}
The functions ${\cal T}^{(n)}
$ involved in expansion (\ref{expT}) are analytic in $\bx_{\sA}$ and $\bx_{\sB}$ when both the following conditions are met:

a) $\bx_{\sB}\neq\bx_{\sA}$;

b) $\bn_{\sB}\neq -\bn_{\sA}$.
\end{proposition}

The importance of this property will clearly appear in section \ref{princmeth}. It is worth of noting that the second condition in proposition \ref{analytT} is automatically fulfilled when inequality (\ref{strl}) is satisfied. This fact explains why the condition {\it b)} is never explicitly involved in the assumptions of the propositions enunciated below.

\section{First procedure: determination of the ${\cal T}^{(n)}$'s from the recurrence relation for $n=1, 2, 3$} \label{Determ1}

The recurrence relation yielded by proposition \ref{itTAB} enables us to determine explicitly the functions ${\cal T}^{(n)}$ at least up to the third order. The calculations are made easier by using the expressions of the light direction triples up to the order $G^2$ performed in \cite{teyssandier2}. Indeed, substituting for ${\cal T}$ from (\ref{expT}) into (\ref{hle})-(\ref{hlr}), it may be seen that the vector functions $\widehat{\underline{\bl}}_{\, e}$ and $\widehat{\underline{\bl}}_{\, r}$ can be expanded in power series of $G$ as follows
\begin{eqnarray}
& &\widehat{\underline{\bl}}_{\, e}(\bx_{\sA}, \bx_{\sB})=-\bN_{\sA\sB}+\sum_{n=1}^{\infty}\widehat{\underline{\bl}}_{\, e}^{(n)}(\bx_{\sA}, \bx_{\sB}), \label{exphle} \\
& &\widehat{\underline{\bl}}_{\, r}(\bx_{\sA}, \bx_{\sB})= -\bN_{\sA\sB}+\sum_{n=1}^{\infty}\widehat{\underline{\bl}}_{\, r}^{(n)}(\bx_{\sA}, \bx_{\sB}), \label{exphlr}
\end{eqnarray}
where the contributions of order $n$ are determined by 
\begin{eqnarray}
& &\widehat{\underline{\bl}}_{\, e}^{(n)}(\bx_{\sA}, \bx_{\sB})=c\bm \nabla_{\bx_{\sA}}{\cal T}^{(n)}(\bx_{\sA}, \bx_{\sB}), \label{hlen} \\
& &\widehat{\underline{\bl}}_{\, r}^{(n)}(\bx_{\sA}, \bx_{\sB})=-c\bm \nabla_{\bx_{\sB}}{\cal T}^{(n)}(\bx_{\sA}, \bx_{\sB}) . \label{hlrn}
\end{eqnarray}
As a consequence, the recurrence relation (\ref{FABn}) may be written in the form 
\begin{eqnarray}
& &\hspace{-10mm}F^{(n)}(\bx_{\sA},\bx_{\sB})=\kappa_n m^n \int_{0}^{1}\frac{d\xi}{\vert \bz(\xi)\vert^n} \nonumber \\
& &\quad\quad-\frac{1}{2}\int_{0}^{1}\sum_{p=1}^{n-1}\left[\widehat{\underline{\bl}}_{\, r}^{(p)}(\bx_{\sA}, \bz(\xi)) . \widehat{\underline{\bl}}_{\, r}^{(n-p)}(\bx_{\sA}, \bz(\xi))\right] d\xi. \label{FABn1}
\end{eqnarray}
whatever $n\geq 2$. The results inferred from (\ref{TnFn}), (\ref{FAB1}) and (\ref{FABn1}) for $n=1, 2, 3$ may be enunciated as follows.

\begin{proposition} \label{propT3}
Let $\bx_{\sA}$ and $\bx_{\sB}$ be two points in ${\cal D}_h$ such that both the conditions $\bn_{\sA}\neq\bn_{\sB}$ and (\ref{strl}) are met. For $n=1, 2, 3$, the functions ${\cal T}^{(n)}$ are yielded by
\begin{eqnarray} 
& &\hspace{-2mm}{\cal T}^{(1)}( \bx_{\sA}, \bx_{\sB}) = \frac{(1+\gamma)m}{c}\ln \left(\frac{r_{\sA}+r_{\sB}+\vert \bx_{\sB} -\bx_{\sA} \vert}{r_{\sA}+r_{\sB}-\vert \bx_{\sB} -\bx_{\sA} \vert}\right), \label{T1} \\
& &\hspace{-2mm}{\cal T}^{(2)}( \bx_{\sA}, \bx_{\sB}) = \frac{m^2}{r_{\sA}r_{\sB}} \frac{\vert \bx_{\sB} -\bx_{\sA} \vert}{c}\bigg[ \kappa\frac{\arccos \bn_{\sA}.\bn_{\sB}}{\vert\bn_{\sA}\times\bn_{\sB}\vert}-
\frac{(1+\gamma)^2}{1+\bn_{\sA}.\bn_{\sB}}\bigg], \label{T2} \\
& &\hspace{-2mm}{\cal T}^{(3)}(\bx_{\sA},\bx_{\sB})=\frac{m^3}{r_{\sA}r_{\sB}}\left( \frac{1}{r_{\sA}}+\frac{1}{r_{\sB}}\right)\frac{\vert\bx_{\sB}-\bx_{\sA}\vert}{c(1+\bn_{\sA}.\bn_{\sB})} \bigg\lbrack \kappa_3 -(1+\gamma)\kappa\frac{\arccos \bn_{\sA} .\bn_{\sB}}{\vert\bn_{\sA}\times\bn_{\sB}\vert}  \nonumber \\
& &\qquad\qquad\qquad\;\;\;+\frac{(1+\gamma )^3}{1+\bn_{\sA}.\bn_{\sB}}\bigg\rbrack, \label{T3}
\end{eqnarray}
where the coefficients $\kappa$ and $\kappa_3$ are determined by (\ref{kappa}) and (\ref{kappa3}), respectively. In general relativity, $\gamma$, $\kappa$ and $\kappa_3$ are given by
\be \label{GRk}
\gamma =1,\quad\kappa=\mbox{$\frac{15}{4}$}, \quad \kappa_3=\mbox{$\frac92$}.
\ee
\end{proposition}

Before entering the proof of this proposition, it is worthy of note that equations (\ref{Tr1})-(\ref{Tr3}) corresponding to a radial light ray are recovered by taking the limit of equations (\ref{T1})-(\ref{T3}) when $\bn_{\sB}\rightarrow \bn_{\sA}$. One has indeed 
\be \label{lim2}
\lim_{\bn_{\sB}\rightarrow \bn_{\sA}}\frac{\arccos \bn_{\sA}.\bn_{\sB}}{\vert\bn_{\sA}\times\bn_{\sB}\vert}=1.
\ee 

Expression (\ref{T1}) is straightforwardly obtained from (\ref{FAB1}) under the form
\be \label{Shap}
{\cal T}^{(1)}(\bx_{\sA}, \bx_{\sB})=\frac{(1+\gamma)m}{c}\ln\left(\frac{r_{\sB}+\bN_{\sAB}.\bx_{\sB}}{r_{\sA}+\bN_{\sAB}.\bx_{\sA}}\right),
\ee
which coincides with the well-known Shapiro time delay \index{Shapiro time delay} expressed in a standard post-Newtonian gauge (see, e.g., \cite{will}). The equivalent formula given by (\ref{T1}) is more convenient for deriving the first-order light direction triples (see, e.g., \cite{blanchet} and \cite{linet}).

Equation (\ref{T2})  has  been obtained in \cite{leponcin1} and \cite{teyssandier1} (see also \cite{klioner1} for an equivalent expression in an harmonic gauge). Nevertheless, we give a detailed proof of proposition \ref{propT3} also for $n=2$ because our calculation is based on a new procedure which in principle can be efficient at any order. 

{\bf Proof of proposition \ref{propT3} for $n=2$ and $n=3$}. For calculating $F^{(2)}$, we just need the triple $\widehat{\underline{\bl}}_{\, r}^{(1)}(\bx_{\sA}, \bx_{\sB})$, which is easily deduced from (\ref{hlrn}) and (\ref{T1}). One has (see, e.g., \cite{teyssandier2}):
\be \label{l1r}
\widehat{\underline{\bl}}_{\, r}^{(1)}(\bx_{\sA}, \bx_{\sB})=-\frac{(1+\gamma)m\vert \bN_{\sAB}\times\bn_{\sB}\vert}{r_c}\left[\bN_{\sAB}-\frac{\vert \bn_{\sA}\times \bn_{\sB}\vert}{1+\bn_{\sA} . \bn_{\sB}}\bP_{\sAB}\right],
\ee
where $\bm P_{\sA\sB}$ is defined as
\be \label{PAB}
\bm P_{\sAB}=\bN_{\sAB}\times\left(\frac{\bm n_{\sA}\times\bm n_{\sB}}{\vert\bm n_{\sA}\times\bm n_{\sB}\vert}\right).
\ee
Substituting $\bz(\xi)$ for $\bx_{\sB}$ in (\ref{l1r}), defining $\bn(\xi)$ as
\be \label{nxi} 
\bn(\xi)=\frac{\bz(\xi)}{\vert\bz(\xi)\vert},
\ee
and then pointing out that 
\be \label{Nxi}
\frac{\bz(\xi)-\bx_{\sA}}{\vert \bz(\xi)-\bx_{\sA} \vert}=\bN_{\sAB},
\ee
it may be seen that relation (\ref{FABn1}) reduces to
\be \label{FAB2b}
F^{(2)}(\bx_{\sA},\bx_{\sB})=\kappa m^2 \int_{0}^{1}\frac{d\xi}{\vert \bz(\xi)\vert^2}-(1+\gamma)^2 m^2\int_{0}^{1}\frac{1}{1+\bn_{\sA}.\bn(\xi)}\frac{1}{\vert\bz(\xi)\vert^2} d\xi
\ee
when $n=2$. The first integral in (\ref{FAB2b}) is elementary. The second one is easily calculated using a relation already exploited in \cite{teyssandier1}, namely
\[
\frac{1}{1+\bn_{\sA}.\bn(\xi)}\frac{1}{\vert\bz(\xi)\vert^2}=\frac{d}{d\xi}\left[\frac{\xi}{r_{\sA}\vert\bz(\xi)\vert+\bx_{\sA}.\bz(\xi)}\right].
\]
However, finding such a  procedure for the higher-order terms cannot be reasonably expected. So we propose in this section another method based on a change of variable which appreciably simplifies the calculations and may be succesfully applied at least up to the third order. 

Since the metric is spherically symmetric, a non-radial light ray \index{light ray} $\Gamma_{s}(\bx_{\sA},\bx_{\sB})$ is confined to the plane defined by $\bx_{\sA}$ and $\bx_{\sB}$. We assume that this plane is the equatorial plane defined by $\vartheta=\pi/2$. The parameter $\xi$ is a monotonic function of the angular coordinate $\varphi$. As a consequence, $\varphi$ can be used as a variable of integration in formulae (\ref{FAB1}) and (\ref{FABn}).  For the sake of brevity, we assume that the direction of the light propagation is such that $\varphi-\varphi_{\sA}>0$ during the motion of the photon. Let us denote by $H$ the foot of the perpendicular drawn from the origin $O$ of the spatial coordinates to the straight line passing through $\bx_{\sA}$ and $\bx_{\sB}$. If $\varphi_c$ is the value of $\varphi$ for $H$, an elementary geometric reasoning shows that 
\be \label{iz}
\vert \bz(\xi)\vert=\frac{r_c}{\cos(\varphi-\varphi_c)}
\ee 
and 
\be \label{dxi}
d\xi = \frac{\vert \bz(\xi)\vert^2}{r_c\vert \bx_{\sB}-\bx_{\sA}\vert}\,d\varphi,
\ee
where $r_c$ is the zeroth-order distance of closest approach of the light ray to $O$, namely 
\be \label{rc}
r_c=OH=\frac{r_{\sA}r_{\sB}}{\vert\bx_{\sB}-\bx_{\sA} \vert} \vert \bn_{\sA}\times \bn_{\sB} \vert.
\ee

Taking into account the relation 
\be \label{cos}
\bn_{\sA}.\bn(\xi)=\cos(\varphi-\varphi_{\sA}),
\ee
it appears that $F^{(2)}$ may be written in the form
\be \label{F2inta}
F^{(2)}(\bx_{\sA},\bx_{\sB}) =\frac{m^2}{r_c\vert \bx_{\sB}-\bx_{\sA}\vert} [\kappa\Phi_{12}-(1+\gamma)^2\Phi_{22}],
\ee
where 
\be \label{Phi2}
\Phi_{12}=\int_{\varphi_{\sA}}^{\varphi_{\sB}}d\varphi, \qquad \Phi_{22}=\int_{\varphi_{\sA}}^{\varphi_{\sB}}\frac{d\varphi}{1+\cos(\varphi-\varphi_{\sA})}.
\ee

Integrals $\Phi_{12}$ and $\Phi_{22}$ are elementary. Finding expressions of these integrals in terms of $r_{\sA}$, $r_{\sB}$, $\bn_{\sA}$ and $\bn_{\sB}$ is easily performed by supplementing (\ref{cos}) with a relation as follows
\be \label{sin}
\vert\bn_{\sA}\times\bn(\xi)\vert=\sin(\varphi-\varphi_{\sA}).
\ee
We get
\be \label{Phi3}
\Phi_{12}=\arccos \bn_{\sA}.\bn_{\sB},  \qquad\Phi_{22}= \frac{\vert\bn_{\sA}\times\bn_{\sB}\vert}{1+\bn_{\sA}.\bn_{\sB}}.
\ee 
Hence expression (\ref{T2}) for ${\cal T}^{(2)}$.

The same procedure may be applied for determining $F^{(3)}$. Substituting for ${\cal T}^{(2)}$ from (\ref{T2}) in (\ref{hlrn}) yields (see \cite{teyssandier2})
\begin{eqnarray}
&&\hspace{-15mm}\widehat{\underline{\bl}}_{\, r}^{(2)}(\bx_{\sA}, \bx_{\sB})=-\frac{m^2 \vert\bN_{\sAB}\times\bn_{\sB}\vert}{r_c^2}\bigg\lbrace \vert\bN_{\sAB}\times\bn_{\sB}\vert\left[\kappa - \frac{(1+\gamma)^2}{1+\bn_{\sA} . \bn_{\sB}}\right]\bN_{\sAB} \nonumber \\
&&\qquad\; +\bigg\lbrace \kappa\bigg\lbrack\frac{\arccos\bn_{\sA} . \bn_{\sB}}{\vert\bn_{\sA}\times\bn_{\sB}\vert} \bN_{\sAB} . \bn_{\sA} -\bN_{\sAB} . \bn_{\sB}\bigg\rbrack \nonumber \\
&&\qquad\; +(1+\gamma)^2\frac{\bN_{\sAB} . \bn_{\sB}-\bN_{\sAB} . \bn_{\sA}}{1+\bn_{\sA} . \bn_{\sB}}\bigg\rbrace\bP_{\sAB} \bigg\rbrace .\label{l2r}
\end{eqnarray}
Then, calculating $\widehat{\underline{\bl}}_{\, r}^{(1)} . \widehat{\underline{\bl}}_{\, r}^{(2)}$ from (\ref{l1r}) and (\ref{l2r}), it is easily deduced from (\ref{FABn1}) that $F^{(3)}$ may be written in the form
\be \label{T3inta}
F^{(3)}(\bx_{\sA},\bx_{\sB}) = \frac{m^3}{r_c^2 \vert \bx_{\sB} - \bx_{\sA}\vert}\left[\kappa_3\Phi_{13} - (1+\gamma)\kappa\,\Phi_{23} + (1+\gamma)^3\Phi_{33}\right],
\ee
where $\Phi_{13} , \Phi_{23}$ and $\Phi_{33}$ are defined by 
\begin{eqnarray}
&&\hspace{-2mm}\Phi_{13} = r_c^2\vert \bx_{\sB} - \bx_{\sA}\vert\int^{1}_{0}\frac{d\xi}{\vert\bz(\xi)\vert^3}, \label{Phi13int0}\\
&&\hspace{-2mm}\Phi_{23} = \frac{\vert \bx_{\sB} - \bx_{\sA}\vert}{r_c}\int^{1}_{0}\bigg\lbrack\vert\bN_{\sAB}\times\bn(\xi)\vert+\frac{\vert\bn_{\sA}\times\bn(\xi)\vert(\bN_{\sAB}.\bn(\xi))}{1+\bn_{\sA}.\bn(\xi)} \nonumber \\
&&\qquad\qquad-(\bN_{\sAB}.\bn_{\sA})\frac{\arccos\bn_{\sA}.\bn(\xi)}{1+\bn_{\sA}.\bn(\xi)}\bigg\rbrack \vert\bN_{\sAB}\times\bn(\xi)\vert^2 d\xi,     \label{Phi23int0} \\
&&\hspace{-2mm}\Phi_{33} = \frac{\vert \bx_{\sB} - \bx_{\sA}\vert}{r_c}\int^{1}_{0}\bigg\lbrace\frac{\vert\bN_{\sAB}\times\bn(\xi)\vert}{1+\bn_{\sA}.\bn(\xi)}\nonumber \\
&&\qquad\qquad+\vert\bn_{\sA}\times\bn(\xi)\vert\frac{\bN_{\sAB}.\bn(\xi)-\bN_{\sAB}.\bn_{\sA}}{[1+\bn_{\sA}.\bn(\xi)]^2}\bigg\rbrace\vert\bN_{\sAB}\times\bn(\xi)\vert^2 d\xi. \label{Phi33int0}
\end{eqnarray}
Using (\ref{iz})-(\ref{dxi}), (\ref{cos}) and (\ref{sin}) supplemented with
\be \label{cosi}
\vert\bN_{\sAB}\times\bn(\xi)\vert=\cos(\varphi-\varphi_c), \qquad\bN_{\sAB}.\bn(\xi)=\sin(\varphi-\varphi_c),
\ee
$\Phi_{13} , \Phi_{23}$ and $\Phi_{33}$ may be rewritten in the form
\begin{eqnarray}
&&\hspace{-8mm}\Phi_{13} = \int_{\varphi_{\sA}}^{\varphi_{\sB}}\cos(\varphi - \varphi_{c})d\varphi, \label{Phi13int}\\
&&\hspace{-8mm}\Phi_{23} = \int_{\varphi_{\sA}}^{\varphi_{\sB}}\left[\cos(\varphi - \varphi_{c})+\frac{\sin(\varphi-\varphi_{\sA}) \sin(\varphi - \varphi_{c})}{1+\cos(\varphi-\varphi_{\sA})} \right.\nonumber \\
&&\;\;\left.-\sin(\varphi_{\sA} - \varphi_{c})\frac{\varphi-\varphi_{\sA}}{1+\cos(\varphi-\varphi_{\sA})}\right]d\varphi,     \label{Phi23int} \\
&&\hspace{-8mm}\Phi_{33} = \int_{\varphi_{\sA}}^{\varphi_{\sB}}\left\{\frac{\cos(\varphi - \varphi_{c})}{1+\cos(\varphi-\varphi_{\sA})}\right. \nonumber \\
&&\;\;\left.+\sin(\varphi-\varphi_{\sA}) \frac{\sin(\varphi - \varphi_{c}) - \sin(\varphi_{\sA} - \varphi_{c})}{[1+\cos(\varphi-\varphi_{\sA})]^2}\right\}d\varphi. \label{Phi33int}
\end{eqnarray}
Noting that $\varphi - \varphi_c = \varphi -\varphi_{\sA}+ \varphi_{\sA} - \varphi_{c}$, and then using the trigonometric formulae developing the sine and cosine of a sum of angles, it may be seen that (\ref{Phi23int}) and (\ref{Phi33int}) transform into 
\begin{eqnarray}
&&\Phi_{23}=-\sin(\varphi_{\sA} - \varphi_{c})\int_{\varphi_{\sA}}^{\varphi_{\sB}}\frac{\varphi-\varphi_{\sA}+\sin(\varphi-\varphi_{\sA})}{1+\cos(\varphi-\varphi_{\sA})} d\varphi  \nonumber \\
&&\qquad\quad+(\varphi_{\sB}-\varphi_{\sA})\cos(\varphi_{\sA} - \varphi_{c}), \label{Ph23ib}\\
&&\medskip \nonumber \\
&&\Phi_{33} = \int_{\varphi_{\sA}}^{\varphi_{\sB}}\left\{\frac{\cos(\varphi_{\sA} - \varphi_{c})}{1+\cos(\varphi-\varphi_{\sA})}-2\frac{\sin(\varphi_{\sA} - \varphi_{c})\sin(\varphi-\varphi_{\sA})}{[1+\cos(\varphi-\varphi_{\sA})]^2}\right\}d\varphi.\nonumber \\
&& \label{Ph33ib}
\end{eqnarray}
Integrating expressions (\ref{Ph23ib}) and (\ref{Ph33ib}) is straightforward. Taking into account (\ref{cosi}) written for $\xi=1$ and noting that 
\be \label{NABnn}
\vert\bN_{\sAB}\times\bn_{\sA}\vert=\frac{r_{\sB}\vert\bn_{\sA}\times\bn_{\sB}\vert}{\vert \bx_{\sB} - \bx_{\sA}\vert},\qquad\vert\bN_{\sAB}\times\bn_{\sB}\vert=\frac{r_{\sA}\vert\bn_{\sA}\times\bn_{\sB}\vert}{\vert \bx_{\sB} - \bx_{\sA}\vert},
\ee
we get
\begin{eqnarray}
&&\Phi_{13} = \frac{r_{\sA}+r_{\sB}}{\vert \bx_{\sB} - \bx_{\sA}\vert}(1 - \bn_{\sA} . \bn_{\sB}), \nonumber \\
&&\Phi_{23} = \Phi_{13} \frac{\arccos\bn_{\sA}.\bn_{\sB}}{\vert \bn_{\sA} \times \bn_{\sB}\vert}, \nonumber \\
&&\Phi_{33} = \Phi_{13} \frac{1}{1 + \bn_{\sA} . \bn_{\sB}}. \nonumber
\end{eqnarray}
Hence equation (\ref{T3}) for ${\cal T}^{(3)}$. 

The results of this section show that proposition \ref{itTAB} enables us to perform the calculation of ${\cal T}^{(1)}$, ${\cal T}^{(2)}$ and ${\cal T}^{(3)}$. It is probable that the recurrence relation (\ref{FABn}) allows explicit calculations for $n\geq4$. However, it is to be feared that unwieldy calculations have to be performed. So we set out another procedure, recently proposed in \cite{linet2}. As it has been emphasized in the introduction, this procedure only involves elementary integrations whatever the order of approximation.

\section{Second procedure: determination of the ${\cal T}^{(n)}$'s from the geodesic equations} \label{Determ2}

\subsection{Null geodesic equations \index{null geodesic equations}} \label{Ngeqns}

Let $\Gamma$ be an arbitrary non-radial null geodesic \index{null geodesic} path of the metric $d\tilde{s}^2$. We suppose that $\Gamma$ is confined in the region ${\cal D}_h$ and described by parametric equations $x^{\alpha}=x^{\alpha}(\zeta)$, where $\zeta$ is an arbitrarily chosen affine parameter. We choose again the spherical coordinates $(r,\vartheta ,\varphi )$ so that $\vartheta = \pi /2$ for any point of this path. Denoting by $\tilde{l}_{\alpha}$ the covariant components of the vector tangent to $\Gamma_{s}(\bx_{\sA},\bx_{\sB})$, an equation as follows 
\be \label{ldx0}
\tilde{l}_0 dx^0+\tilde{l}_r dr+\tilde{l}_{\varphi}d\varphi=0
\ee
is satisfied along $\Gamma$ since $\tilde{l}_{\alpha}$ is a null vector. Owing to the symmetries of the metric, we have 
\begin{eqnarray} 
&&\tilde{l}_0=E, \label{ip1} \\
&&\tilde{l}_{\varphi}=-J, \label{ip2} 
\end{eqnarray}
with $E$ and $J$ being constants of the motion. For convenience, the affine parameter $\zeta$ is chosen in such a way that $E>0$. Furthermore, it is always possible to suppose $J>0$ without lack of generality when calculating the time transfer function in a static,  \index{spherically symmetric spacetime} spherically symmetric spacetime. Then the quantity defined as
\be \label{db}
b=\frac{J}{E}
\ee
is the impact parameter of the light ray \index{impact parameter of a light ray} (see, e.g., \cite{chandra} and \cite{teyssandier2})\footnote{$b$ is an intrinsic quantity attached to $\Gamma$ since the constants of the motion $E$ and $J$ are themselves coordinate-independent quantities.}. It may be noted that $b=0$ would correspond to a radial null geodesic.

Since $d\tilde{s}^2=0$ along $\Gamma$, it follows from (\ref{ip1}), (\ref{ip2}) and (\ref{db}) that
\be \label{eqdf2}
\tilde{l}_r=-\varepsilon \frac{E}{r}\sqrt{r^2{\cal U}(r)-b^2},
\ee
where $\varepsilon=1$ when $r$ is an increasing function of time and $\varepsilon=-1$ when $r$ is a decreasing function of time\footnote{The sign of $\varepsilon$ in equation (\ref{eqdf2}) is changed if and only if the photon passes through a pericenter or an apocenter. The passage through an apocenter corresponds to an extreme relativistic case.}. Substituting for $\tilde{l}_r$ from (\ref{eqdf2}) into (\ref{ldx0}), and then dividing throughout by $E$, we get a relation enabling to determine the light travel time \index{light travel time} by an integration along $\Gamma$, namely
\be \label{dx0a}
dx^0 = b d\varphi+\frac{\varepsilon}{r}\sqrt{r^2{\cal U}(r)-b^2} dr.
\ee
 
Let us assume now that $\Gamma$ is a quasi-Minkowskian light ray $\Gamma_s(\bx_{\sA }, \bx_{\sB })$. As we shall see in what follows, our procedure for calculating explicitly the corresponding perturbation functions ${\cal T}^{(n)}$ rests on the property shown in the next subsection that the impact parameter $b$ can be determined as a function of $\bx_{\sA}$ and $\bx_{\sB}$ under the form of an expansion in a series in powers of $G$ by taking the partial derivative of ${\cal T}$ with respect to the cosine of the angle formed by $\bx_{\sA}$ and $\bx_{\sB}$.

\subsection{Post-Minkowskian expansion of the impact parameter} \label{const_integr}

Let $[\varphi_{\sA}, \varphi_{\sB}]$ be the range of the angular function $\varphi (t)$ along a quasi-Minkowskian light ray $\Gamma_{s}(\bx_{\sA},\bx_{\sB})$. For the sake of brevity, we shall frequently use a notation as follows 

\be \label{mu}
\mu=\bn_{\sA}.\bn_{\sB}=\cos(\varphi_{\sB}-\varphi_{\sA}).
\ee
Using this notation, the time transfer function may be considered as a function of $r_{\sA}, r_{\sB}$ and $\mu$:
\[
{\cal T}(\bx_{\sA},\bx_{\sB})={\cal T}(r_{\sA},r_{\sB},\mu).
\]
It is then possible to enunciate the following proposition.

\begin{proposition} \label{bexpm}
Let $\bx_{\sA}$ and $\bx_{\sB}$ be two points in ${\cal D}_h$ such that both the conditions $\bn_{\sA} \neq \bn_{\sB}$ and (\ref{strl}) are fulfilled. The impact parameter \index{impact parameter of a light ray} $b$ of a quasi-Minkowskian light ray joining $\bx_{\sA}$ and $\bx_{\sB}$ may be expanded in powers of $G$ as follows:
\be \label{expb}
b=r_c\left[ 1+\sum_{n=1}^{\infty}\left( \frac{m}{r_c}\right)^n q_n\right],
\ee
where $r_c$ is defined by (\ref{rc}) and the quantities $q_n$ are given by
\be \label{qn}
q_n=-c\left(\frac{r_{c}}{m}\right)^n\frac{\sqrt{1-\mu^2}}{r_c}\,\frac{\partial {\cal T}^{(n)}(r_{\sA},r_{\sB},\mu)}{\partial \mu}.
\ee
\end{proposition}

{\bf Proof of proposition \ref{bexpm}.}  Noting that
\[
\vert\bn_{\sA}\times\bn_{\sB}\vert = \sqrt{1-\mu^2},
\]
it is immediately inferred from equation (13) in \cite{teyssandier2} that the impact parameter of $\Gamma_{s}(\bx_{\sA}, \bx_{\sB})$ may be rewritten in the form
\be \label{bdT}
b=-c\sqrt{1-\mu^2} \, \frac{\partial {\cal T}(r_{\sA},r_{\sB},\mu)}{\partial \mu}.
\ee  
Substituting for ${\cal T}$ from (\ref{expT}) into (\ref{bdT}) directly leads to the expansion given by (\ref{expb}). The zeroth-order term is easily derived from the elementary formula
\be \label{R}
\vert \bx_{\sB}-\bx_{\sA} \vert =\sqrt{r_{\sA}^{2}-2r_{\sA}r_{\sB}\mu +r_{\sB}^{2}}. 
\ee
Indeed, using (\ref{R}) and taking (\ref{rc}) into account yield 
\be \label{derR}
\frac{\partial \vert \bx_{\sB} - \bx_{\sA}\vert}{\partial  \mu}=-\frac{r_c}{\sqrt{1-\mu^2}}.
\ee

We shall see in the next section that the expression of the time transfer function corresponding to a quasi-Minkowskian light ray can be straightforwardly deduced from proposition \ref{bexpm}.

\subsection{Implementation of the method} \label{princmeth}

If $\Gamma_{s}(\bx_{\sA},\bx_{\sB})$ passes through a pericenter $\bx_{\scriptscriptstyle P}$, the integration of (\ref{dx0a}) requires the determination of $\vert\bx_{\scriptscriptstyle P}\vert$ as a function of $\bx_{\sA}$ and $\bx_{\sB}$. The calculation of the time transfer function is very complicated for such a configuration. Fortunately, owing to the analytic extension theorem, it follows from proposition \ref{analytT} that it is sufficient to determine the expression of each term ${\cal T}^{(n)}$ as a function of $\bx_{\sA}$ and $\bx_{\sB}$ in an arbitrarily chosen open subset of the domain of analyticity. For this reason, the calculation of the ${\cal T}^{(n)}$ are henceforth carried out under the assumption that $\bx_{\sA}$ and $\bx_{\sB}$ fulfil the following conditions:

{\it a)} The radial variable $r$ along a quasi-Minkowskian null geodesic \index{null geodesic} joining $\bx_{\sA}$ and $\bx_{\sB}$ is an increasing function of $t$:
\be \label{rinc}
\frac{dr}{dt} >0, \qquad t_{\sA}\leq t \leq t_{\sB}.
\ee

{\it b)} An inequality as follows
\be \label{cdc}
\bN_{\!\sAB} . \bn_{\sA} > 0
\ee
is satisfied, with $\bN_{\!\sAB}$ being defined by (\ref{NAB}).

These conditions considerably simplify the calculations. Indeed, (\ref{rinc}) eliminates the occurrence of any pericenter (or apocenter) between the emission and the reception of light and (\ref{cdc}) implies that the projection $H$ of the origin $O$ on the straight line passing through $\bx_{\sA}$ and $\bx_{\sB}$ lies outside the straight segment linking $\bx_{\sA}$ and $\bx_{\sB}$. One has therefore 
\be \label{rcrAB}
r_c <r_{\sA}\leq r \leq r_{\sB}
\ee
for any point of $\Gamma_{s}(\bx_{\sA},\bx_{\sB})$. These inequalities ensure that condition (\ref{strl}) is met, since $r_h<r_{\sA}$ for any point $\bx_{\sA}$ located in ${\cal D}_{h}$. 

Under these assumptions, integrating (\ref{dx0a}) along $\Gamma_{s}(\bx_{\sA},\bx_{\sB})$ is straightforward since the range of the angular function $\varphi(r)$ between the emission and the reception of the photon is given by
\be \label{phiAB}
\varphi_{\sB}-\varphi_{\sA}=\arccos\mu.
\ee
Noting that in this case $\varepsilon=1$, it may be seen that the time transfer function \index{time transfer function} is then related to the impact parameter \index{impact parameter of a light ray} $b$ by an equation as follows
\be \label{tab}
{\cal T}(\bx_{\sA},\bx_{\sB})=\frac{1}{c}\left[b\arccos\mu+\int_{r_{\sA}}^{r_{\sB}}\frac{1}{r}{\sqrt{r^2{\cal U}(r)-b^2}}dr\right].
\ee

Since $b$ is a function of $\bx_{\sA}$ and $\bx_{\sB}$ determined by (\ref{bdT}), (\ref{tab}) has to be regarded as an integro-differential equation satisfied by ${\cal T}$. In order to solve this integro-differential equation by an iterative procedure, let us substitute (\ref{invAB}) for ${\cal U}$ and (\ref{expb}) for $b$. Expanding $\sqrt{r^2{\cal U}(r)-b^2}/r$ in a power series in $m/r_c$, rearranging the terms and introducing the notation
\be \label{s}
s=\sqrt{r^2-r_{c}^{2}},
\ee
we get an expression as follows for ${\cal T}$
\begin{eqnarray}
& &\hspace{-10mm}{\cal T}(\bx_{\sA},\bx_{\sB})=\frac{1}{c}\left[r_c\arccos\mu+\int_{r_{\sA}}^{r_{\sB}}\frac{s}{r}\, dr\right] \nonumber \\
& &+\frac{1}{c}\sum_{n=1}^{\infty}\left(\frac{m}{r_c}\right)^n\Bigg\lbrace r_c q_n\arccos\mu
+\int_{r_A}^{r_B}\left[U_n-\frac{r_c^2q_n}{rs}\right]dr\Bigg\rbrace, \label{tABi}
\end{eqnarray}
where each $U_n$ is a function of $r$ which may be written in the form
\begin{eqnarray}  
&&U_1=\frac{(1+\gamma)r_c}{s}, \label{U1} \\
&&U_n=\sum_{k=0}^{3n-4}U_{kn}(q_1,\dots ,q_{n-1})r_c^{3n-k-2}
\frac{r^{k-n+1}}{s^{2n-1}} \label{Xn}
\end{eqnarray}
for $n\geq2$, with the quantities $U_{kn}(q_1,\dots ,q_{n-1})$ being polynomials in $q_1,\dots,q_{n-1}$. 
Noting that
\be \label{RAB1}
r_c\arccos\mu+\int_{r_{\sA}}^{r_{\sB}}\frac{s}{r} \, dr=\vert \bx_{\sB}-\bx_{\sA} \vert
\ee
and  
\be \label{fab0}
\arccos\mu -r_c\int_{r_{\sA}}^{r_{\sB}}\frac{dr}{rs}=0
\ee
when conditions (\ref{rinc}) and (\ref{cdc}) are met\footnote{Note that (\ref{RAB1}) is just (\ref{tab}) written in the case where the gravitational field vanishes, i.e., $m=0$.}, (\ref{expT}) is immediately recovered from (\ref{tABi}), with each perturbation term being given by
\be \label{TnAB}
{\cal T}^{(n)}(\bx_{\sA},\bx_{\sB})=\frac{1}{c}\left(\frac{m}{r_c}\right)^n\int_{r_A}^{r_B}U_n dr.
\ee

As it has been explained in the beginning of this subsection, the expression of ${\cal T}^{(n)}$ as a function of $\bx_{\sA}$ and $\bx_{\sB}$ derived from (\ref{TnAB}) can be regarded as valid even when conditions (\ref{rinc}) and (\ref{cdc}) are not met. In this sense, (\ref{TnAB}) constitutes the main ingredient of the procedure developed in the present section. 

The fact that the coefficient $q_n$ is not involved in $U_n$ and the property for each coefficient $q_k$ to be proportional to a derivative of the function ${\cal T}^{(k)}$ imply that ${\cal T}^{(n)}$ can be determined when the sequence of functions ${\cal T}^{(1)},\dots,{\cal T}^{(n-1)}$ is known. To initiate the process, it is sufficient to infer the expression of ${\cal T}^{(1)}$ from (\ref{U1}) and (\ref{TnAB}). The integration is immediate. Noting that
 \begin{eqnarray} 
\sqrt{r_{\sA}^{2}-r_{c}^{2}}=r_{\sA}\bN_{\sAB}.\bn_{\sA}, \label{NnA} \\
\sqrt{r_{\sB}^{2}-r_{c}^{2}}=r_{\sB}\bN_{\sAB}.\bn_{\sB} \label{NnB}
\end{eqnarray}
when conditions (\ref{rinc}) and (\ref{cdc}) are met, we get again (\ref{Shap}), as it could be expected. Substituting for ${\cal T}^{(1)}$ from (\ref{T1}) into (\ref{qn}) written for $n=1$, and then using (\ref{derR}), it is easily seen that
\be \label{q1f}
q_1=\frac{(1+\gamma) r_c}{1+\bn_{\sA}.\bn_{\sB}}\left( \frac{1}{r_{\sA}}+\frac{1}{r_{\sB}}\right).
\ee
Taking into account this determination of $q_1$, it becomes possible to carry out the calculation of ${\cal T}^{(2)}$ since $U_2$ only involves $q_1$. Then, $q_2$ can be derived from (\ref{qn}) taken for $n=2$. Therefore, ${\cal T}^{(3)}$ can be calculated since $U_3$ only involves $q_1$ and $q_2$, and so on. It may be added that all the integrations involved in the right-hand side of (\ref{TnAB}) are elementary and can be carried out with any symbolic computer program. As a consequence, the procedure set up in this section allows the explicit calculation of ${\cal T}^{(n)}$ as a function of $\bx_{\sA}$ and $\bx_{\sB}$ whatever the order $n$.

\section{Simplification of the second procedure} \label{pract_impl}

Even if it involves only elementary integrals, the procedure developed in the previous section is somewhat tedious. Nevertheless, the method can be notably simplified by making use of the differential equation governing the variation of the angular coordinate along the light ray. 

\subsection{Use of a constraint equation} \label{constr}

Equations (\ref{ip2}) and (\ref{eqdf2}) are equivalent to the geodesic equations
\begin{eqnarray} 
&&\frac{d\varphi}{d\zeta}=\frac{J}{r^2 {\cal U}(r)}, \label{df} \\
&&\frac{dr}{d\zeta}=\varepsilon \frac{E}{r {\cal U}(r)}\sqrt{r^2{\cal U}(r)-b^2} .\label{dr}
\end{eqnarray}
Eliminating the affine parameter $\zeta$ between (\ref{df}) and (\ref{dr}) leads to
\be \label{dfdr}
\frac{d\varphi}{dr}=\varepsilon\frac{b}{r}\frac{1}{\sqrt{r^2{\cal U}(r)-b^2}}.
\ee

Since $\varepsilon=1$ when conditions (\ref{rinc}) and (\ref{cdc}) are met, integrating (\ref{dfdr}) and taking into account (\ref{phiAB}) yield thereby
\be \label{fab}
\arccos\mu=\int_{r_{\sA}}^{r_{\sB}}\frac{b}{r\sqrt{r^2{\cal U}(r)-b^2}}dr.
\ee
Equation (\ref{fab}) may be regarded as a constraining equation which implicitly determines $b$ as a function of $\bx_{\sA}$ and $\bx_{\sB}$. So it may be expected that this equation implies some conditions on the coefficients $q_n$ which may be used to simplify the calculations. 

Replacing ${\cal U}$ by (\ref{invAB}) and $b$ by (\ref{expb}) into (\ref{fab}), it may be seen that  
\be \label{expfab}
\arccos\mu=r_c\int_{r_{\sA}}^{r_{\sB}}\frac{dr}{rs}+\frac{1}{r_c}\sum_{n=1}^{\infty}\left( \frac{m}{r_c}\right)^n \int_{r_A}^{r_B}W_ndr,
\ee
where the $W_n$'s are functions of $r$ which may be written in the form 
\begin{eqnarray}  
&&W_1= -(1+\gamma)\frac{r_c^3 }{s^3}+q_1\frac{r_c^2 r}{s^3}, \label{Y1} \\
&&W_n=\sum_{k=0}^{3(n-1)} W_{kn}(q_1,\dots ,q_{n-1})r_c^{3n-k} 
\frac{r^{k-n+1}}{s^{2n+1}} +q_n\frac{r_c^2 r}{s^3} \label{Yn}
\end{eqnarray}
for $n\geq 2$, with the terms $W_{kn}(q_1,\dots,q_{n-1})$ being polynomials in $q_1,\dots,q_{n-1}$. Taking into account (\ref{fab0}), it is immediately seen that (\ref{expfab}) reduces to
\be \label{expf0}
\sum_{n=1}^{\infty}\left( \frac{m}{r_c}\right)^n \int_{r_A}^{r_B}W_n dr=0
\ee
for $n\geq 1$. Since (\ref{expf0}) holds whatever $m$, it is clear that (\ref{expfab}) is equivalent to the infinite set of equations
\be \label{setY}
\int_{r_{\sA}}^{r_{\sB}}W_n dr=0, \qquad n=1,2,\dots 
\ee

The set of constraint equations (\ref{setY}) may be systematically used for simplifying our problem. Let us consider the functions $U_n^{\ast}$ defined as 
\begin{eqnarray} 
&&U_1^{\ast}=U_1, \label{R1} \\
&&U_n^{\ast}=U_n+\sum_{p=1}^{n-1}k_{pn}W_p \label{Rn}
\end{eqnarray}
for $n\geq 2$, where the $k_{pn}$'s are arbitrary quantities which do not depend on $r$. Taking into account (\ref{setY}), it is immediately seen that
\be \label{XRn}
\int_{r_{\sA}}^{r_{\sB}}U_n dr=\int_{r_{\sA}}^{r_{\sB}}U_n^{\ast}dr.
\ee
Hence ${\cal T}^{(n)}$ may be rewritten in the form
\be \label{tRn}
{\cal T}^{(n)}(\bx_{\sA},\bx_{\sB}) = \frac{1}{c}\left( \frac{m}{r_c}\right)^n \int_{r_{\sA}}^{r_{\sB}}U_n^{\ast} dr .
\ee
Of course, the remark formulated just after (\ref{TnAB}) might be reproduced here. 

It is easily seen that a judicious choice of the quantities $k_{pn}$ enables us to shorten the expressions involved in (\ref{tRn}) when $n\geq2$. Until $n=3$, only the expression of $W_1$ is needed. Indeed, it is easily inferred from the expansion of (\ref{tab}) that $U_2$ and $U_3$ are given by 
\begin{eqnarray} 
& &\hspace{-6mm} U_2=-\frac{\kappa r_c^4}{rs^3}+\frac{(1+\gamma)q_1 r_c^3}{s^3}+\frac{[2\kappa-(1+\gamma)^2-q_1^2]r_c^2r}{2s^3},  \label{U2} \\
& &\hspace{-6mm} U_3=\frac{\kappa_3 r_c^7}{r^2s^5} -\frac{\kappa q_1 r_c^6}{rs^5}- \frac{[2\kappa_3-(1+\gamma)(\kappa +q_1^2-q_2)]r_c^5}{s^5} \nonumber \\
& &\;\;\, +\frac{[2\kappa - 3(1+\gamma)^2-q_1^2+2q_2]q_1r_c^4r}{2s^5} \nonumber \\
& &\;\;\, +\frac{[2\kappa_3-(1+\gamma)(2\kappa - q_1^2-2q_2)+(1+\gamma)^3]r_c^3r^2}{2s^5}-\frac{q_1 q_2r_c^2 r^3}{s^5}.\label{U3}
\end{eqnarray} 
Setting $k_{12}=\frac{1}{2} q_1$ removes the term in $q_1^2$ in $U_2$ and leads to
\be \label{R2}
U_2^{\ast}=-\frac{\kappa r_c^4}{rs^3}+\frac{(1+\gamma)q_1 r_c^3}{2s^3}+\frac{[2\kappa-(1+\gamma)^2]r_c^2r}{2s^3}.
\ee 
Choosing $k_{13}=q_2$ and $k_{23}=0$ remove the terms involving $q_2$ in $U_3$. Then $U_3^{\ast}$ reduces to
\begin{eqnarray} 
& &U_3^{\ast}=\frac{\kappa_3 r_c^7}{r^2s^5} -\frac{\kappa q_1 r_c^6}{rs^5}- \frac{[2\kappa_3-(1+\gamma)(\kappa +q_1^2)]r_c^5}{s^5} \nonumber \\
& &\qquad\;\; +\frac{[2\kappa - 3(1+\gamma)^2-q_1^2]q_1r_c^4r}{2s^5} \nonumber \\
& &\qquad\;\; +\frac{[2\kappa_3-(1+\gamma)(2\kappa - q_1^2)+(1+\gamma)^3]r_c^3r^2}{2s^5}. \label{R3}
\end{eqnarray}
It is thus proved that owing to the constraint equation (\ref{fab}), only the determination of $q_1$ is required for calculating the functions ${\cal T}^{(2)}$ and ${\cal T}^{(3)}$.

{\it Remark}. It may be pointed out that the coefficients $q_n$ could be directly inferred from the constraint equation without differentiating the functions ${\cal T}^{(n)}$ with respect to $\mu$. Indeed, it follows from (\ref{Y1}), (\ref{Yn}) and (\ref{setY}) that
\begin{eqnarray}
& &q_1=\frac{1+\gamma}{r_c}\frac{r_{\sA}\sqrt{r_{\sB}^{2}-r_{c}^{2}}-r_{\sB}\sqrt{r_{\sA}^{2}-r_{c}^{2}}}
{\sqrt{r_{\sB}^{2}-r_{c}^{2}}-\sqrt{r_{\sA}^{2}-r_{c}^{2}}}, \label{q1} \\
& & q_n=-\frac{1}{r_c}\frac{\sqrt{r_{\sA}^{2}-r_{c}^{2}}\,\sqrt{r_{\sB}^{2}-r_{c}^{2}}}{\sqrt{r_{\sB}^{2}-r_{c}^{2}}-\sqrt{r_{\sA}^{2}-r_{c}^{2}}}\nonumber \\
& &\qquad\;\;\times\sum_{k=0}^{3(n-1)} W_{kn}(q_1,\dots ,q_{n-1})r_{c}^{3n-k-1}\int_{r_{\sA}}^{r_{\sB}} 
\frac{r^{k-n+1}}{s^{2n+1}}dr \label{qnc}
\end{eqnarray}
for $n\geq2$. Equation (\ref{qnc}) shows that $q_n$ can be determined once $q_1, \dots,q_{n-1}$ are known. 

It is easily checked that (\ref{q1}) is equivalent to (\ref{q1f}). Indeed, noting that
\be \label{RAB}
\sqrt{r_{\sB}^{2}-r_{c}^{2}}-\sqrt{r_{\sA}^{2}-r_{c}^{2}}=\vert \bx_{\sB}-\bx_{\sA} \vert
\ee
when conditions (\ref{rinc}) and (\ref{cdc}) are met, and then taking into account (\ref{rc}), (\ref{NnA}) and (\ref{NnB}), it may be seen that (\ref{q1}) transforms into
\be \label{q1a}
q_1=(1+\gamma)\frac{\bN_{\sAB}.\bn_{\sB}-\bN_{\sAB}.\bn_{\sA}}{\vert \bn_{\sA}\times\bn_{\sB}\vert}.
\ee
Substituting $r_{\sA}\bn_{\sA}$ for $\bx_{\sA}$ and $r_{\sB}\bn_{\sB}$ for $\bx_{\sB}$ into the numerator of  the right-handside of (\ref{NAB}) yields
\be \label{Nnn}
\bN_{\sAB}.\bn_{\sB}-\bN_{\sAB}.\bn_{\sA}=\frac{(r_{\sA}+r_{\sB})(1-\bn_{\sA}.\bn_{\sB})}{\vert\bx_{\sB}-\bx_{\sA}\vert}.
\ee
Finally, substituting for $\bN_{\sAB}.\bn_{\sB}-\bN_{\sAB}.\bn_{\sA}$ from (\ref{Nnn}) into (\ref{q1a}), and then noting that (\ref{rc}) is equivalent to
\[
\frac{1}{\vert\bx_{\sB}-\bx_{\sA}\vert}=\frac{r_c}{r_{\sA} r_{\sB}} \frac{1}{\vert \bn_{\sA}\times\bn_{\sB}\vert},
\] 
it is immediately seen that (\ref{q1f}) is recovered.

\subsection{Explicit calculation of ${\cal T}^{(1)}$, ${\cal T}^{(2)}$ and ${\cal T}^{(3)}$ } \label{TTF3rd}

We are now in a position to determine the perturbation terms involved in the expansion of the time transfer function up to the order $G^3$. The term ${\cal T}^{(1)}$ has been already treated in section \ref{princmeth}. For $n= 2$ and $n= 3$, it follows from (\ref{R2}) and (\ref{R3}) that ${\cal T}^{(n)}$ may be written in the form
\be \label{intp}
{\cal T}^{(n)}(\bx_{\sA}, \bx_{\sB})=\frac{1}{c}\left(\frac{m}{r_c}\right)^n\sum_{k=0}^{\sigma(n)}U^{\ast}_{kn}(q_1)r_c^{3n-k-2}\int_{r_{\sA}}^{r_{\sB}}\frac{r^{k-n+1}}{s^{2n-1}}\,dr,
\ee
where $\sigma(2)=2$ and $\sigma(3)=4$, with the coefficients $U^{\ast}_{kn}$ being polynomials in $q_1$. The integrals occurring into the right-hand side of (\ref{intp}) are elementary and can be expressed in terms of $r_{\sA}$, $r_{\sB}$, $r_c$, $\sqrt{r_{\sA}^2-r_c^2}$ and $\sqrt{r_{\sB}^2-r_c^2}$. For the explicit calculations, it is convenient to write (\ref{NnA}) and (\ref{NnB}) in the form
\begin{eqnarray} 
\sqrt{r_{\sA}^{2}-r_{c}^{2}}=\frac{r_{\sA}(r_{\sB}\mu -r_{\sA})}{\vert \bx_{\sB} -\bx_{\sA} \vert}, \label{Nn1A} \\ \sqrt{r_{\sB}^{2}-r_{c}^{2}}=\frac{r_{\sB}(r_{\sB} -r_{\sA}\mu)}{\vert \bx_{\sB} -\bx_{\sA} \vert}. \label{Nn1B}
\end{eqnarray}

Using (\ref{rc}), (\ref{q1f}), (\ref{Nn1A}) and (\ref{Nn1B}), it may be seen that ${\cal T}^{(2)}$ and ${\cal T}^{(3)}$ can be expressed in terms of $r_{\sA}r_{\sB}$, $1/r_{\sA}+1/r_{\sB}$, $ \vert \bx_{\sB} -\bx_{\sA} \vert$ and $\mu$. It has been already emphasized that the explicit calculations can be performed with any symbolic computer program. Of course, a simple hand calculation is also possible. For $n=2$ and $n=3$, the calculations are greatly facilitated by noting that (\ref{R2}) and (\ref{R3}) are equivalent to
\be \label{R2b}
\frac{1}{r_c^2}U_2^{\ast}=\frac{\kappa}{rs}+\frac{(1+\gamma)q_1 r_c}{2s^3}-\frac{(1+\gamma)^2r}{2s^3}
\ee
and
\begin{eqnarray} 
& & \frac{1}{r_c^3}U_3^{\ast}=\frac{\kappa_3}{r^2s} -\frac{(1+\gamma) \kappa}{s^3}+ \frac{\kappa q_1r_c}{r s^3}+\frac{(1+\gamma)[(1+\gamma)^2+q_1^2]r^2}{2s^5} \nonumber \\
& &\quad\qquad\quad-\frac{[3(1+\gamma)^2+q_1^2]q_1r_c r}{2s^5}+\frac{(1+\gamma)q_1^2 r_c^2}{s^5}, \label{R3b}
\end{eqnarray}
respectively. Calculating ${\cal T}^{(2)}$ from (\ref{tRn}) and (\ref{R2b}) is elementary and straightforwardly yields (\ref{T2}). Calculating ${\cal T}^{(3)}$ from (\ref{tRn}) and (\ref{R3b}) requires somewhat tedious calculations, which are detailed in an appendix of \cite{linet2}. The result coincides with (\ref{T3}). We have seen in section \ref{princmeth} that the expressions thus obtained can be considered as valid even when conditions (\ref{rinc}) and (\ref{cdc}) are not fulfilled. So we can state that at least up to the third order, the procedures developed in sections \ref{Determ1} and \ref{Determ2} lead to identical expressions for the first three perturbation terms involved in the expansion of the time transfer function. This concordance confirms the reliability of the second procedure presented in this paper.

\section{Direction of light propagation \index{direction of light propagation} up to order $G^3$} \label{imp_par3}

We are now in a position to obtain explicit expressions for the triples giving the direction of light propagation at points $\bx_{\sA}$ and $\bx_{\sB}$ up to the third order in $G$. 
The vector functions $\widehat{\underline{\bl}}_{\, e}^{(n)}$ and $\widehat{\underline{\bl}}_{\, r}^{(n)}$ could be straightforwardly derived for $n=1, 2, 3$ by substituting  for ${\cal T}^{(n)}$ from (\ref{T1})-(\ref{T3}) into (\ref{hlen}) and (\ref{hlrn}). Nevertheless, the calculation is greatly facilitated by making use of formulae (17{\it a}) and (17{\it b}) given in \cite{teyssandier2}. Indeed, taking into account (\ref{expb}), these formulae lead to
\begin{eqnarray}
&&\hspace{-6mm}\widehat{\underline{\bm l}}_{\,e}^{(n)}(\bm x_{\sA},\bm x_{\sB})=\bigg\lbrack c\frac{\partial {\cal T}^{(n)}}{\partial r_{\sA}}\bm N_{\sAB}.\bm n_{\sA}-\frac{m^n}{r_{\sA}^n}\frac{r_{\sA}^{n-1}}{r_c^{n-1}} q_n\vert \bm N_{\sAB}\times\bm n_{\sA}\vert\bigg\rbrack \bm N_{\sAB}  \nonumber \\
&&\qquad\qquad\quad+\bigg\lbrack c\frac{\partial {\cal T}^{(n)}}{\partial r_{\sA}}\vert \bm N_{\sAB}\times\bm n_{\sA}\vert+\frac{m^n}{r_{\sA}^n}\frac{r_{\sA}^{n-1}}{r_c^{n-1}}q_n \bm N_{\sAB}.\bm n_{\sA}\bigg\rbrack \bm P_{\sAB}, \nonumber \\
&&\label{wlA3} \\
&&\hspace{-6mm}\widehat{\underline{\bm l}}_{\,r}(\bm x_{\sA},\bm x_{\sB})=-\bigg\lbrack c\frac{\partial {\cal T}^{(n)}}{\partial r_{\sB}}\bm N_{\sAB}.\bm n_{\sB}+\frac{m^n}{r_{\sB}^n}\frac{r_{\sB}^{n-1}}{r_c^{n-1}} q_n\vert \bm N_{\sAB}\times\bm n_{\sB}\vert\bigg\rbrack \bm N_{\sAB} \nonumber \\
&&\qquad\qquad\quad-\bigg\lbrack c\frac{\partial {\cal T}^{(n)}}{\partial r_{\sB}}\vert \bm N_{\sAB}\times\bm n_{\sB}\vert-\frac{m^n}{r_{\sB}^n}\frac{r_{\sB}^{n-1}}{r_c^{n-1}} q_n \bm N_{\sAB}.\bm n_{\sB}\bigg\rbrack \bm P_{\sAB}, \nonumber \\
&&\label{wlB3}
\end{eqnarray}
where $\bm P_{\sAB}$ is defined by (\ref{PAB}).

Equations (\ref{wlA3}) and (\ref{wlB3}) show that knowing $c\partial {\cal T}^{(n)}/\partial r_{\sA}$, $c\partial {\cal T}^{(n)}/\partial r_{\sB}$ and $q_n$ is sufficient to determine the triples $\widehat{\underline{\bm l}}_{\,e}^{(n)}$ and $\widehat{\underline{\bm l}}_{\,r}^{(n)}$. The coefficient $q_1$ is given by (\ref{q1f}). Replacing $\bn_{\sA}.\bn_{\sB}$ by $\mu$ in (\ref{T2}) and (\ref{T3}), and then taking into account (\ref{R}), $q_2$ and $q_3$ are straightforwardly derived from (\ref{qn}). Noting that
\be \label{rcR}
\frac{r_{\sA}r_{\sB}(1-\mu^2)}{\vert \bx_{\sB}-\bx_{\sA}\vert^2}=\frac{r_c^2}{r_{\sA}r_{\sB}},
\ee
we can formulate the proposition below.

\begin{proposition} \label{propq123}
The coefficients $q_1$, $q_2$ and $q_3$ involved in the expansion of the impact parameter \index{impact parameter of a light ray} $b$ of a quasi-Minkowskian light ray joining $\bx_{\sA}$ and $\bx_{\sB}$ are given by 
\begin{eqnarray} 
& & q_1 (\bx_{\sA} ,\bx_{\sB} )= ( 1+\gamma )\left(\frac{r_c}{r_{\sA}}+\frac{r_c}{r_{\sB}}\right)\frac{1}{1+\bn_{\sA}.\bn_{\sB}},  \label{q1N} \\
& & q_2(\bx_{\sA},\bx_{\sB})=\kappa \left[ 1-\left(\bn_{\sA}.\bn_{\sB}-\frac{r_c^2}{r_{\sA}r_{\sB}}\right)\frac{\arccos \bn_{\sA}.\bn_{\sB}}{\vert\bn_{\sA}\times\bn_{\sB}\vert}\right] \nonumber \\
& &\qquad \qquad\quad \;\; \, -\frac{(1+\gamma )^2}{1+\bn_{\sA}.\bn_{\sB}}\left(1-\bn_{\sA}.\bn_{\sB}+\frac{r_c^2}{r_{\sA}r_{\sB}}\right), \label{q2N}\\
& & q_3 (\bx_{\sA} ,\bx_{\sB})=\left(\frac{r_c}{r_{\sA}}+\frac{r_c}{r_{\sB}}\right)\frac{1}{1+\bn_{\sA}.\bn_{\sB}} \bigg\lbrace\kappa_3 \left(1-\bn_{\sA}.\bn_{\sB}+\frac{r_c^2}{r_{\sA}r_{\sB}}\right)\nonumber \\
& &\qquad \qquad\quad \;\; \, -(1+\gamma )\kappa \left[ 1 +\left(1-2\bn_{\sA}.\bn_{\sB}+\frac{r_c^2}{r_{\sA}r_{\sB}}\right)\frac{\arccos \bn_{\sA}.\bn_{\sB}}{\vert\bn_{\sA}\times\bn_{\sB}\vert} 
\right] \nonumber \\
& &\qquad \qquad\quad \;\; \,+\frac{(1+\gamma )^3}{1+\bn_{\sA}.\bn_{\sB}}\left(2-2\bn_{\sA}.\bn_{\sB}+\frac{r_c^2}{r_{\sA}r_{\sB}}\right) \bigg\rbrace . \label{q3N}
\end{eqnarray}
\end{proposition}

Noting that 
\be \label{iden1}
\frac{1}{r_{\sA}}=\frac{\vert\bN_{\sAB}\times\bn_{\sA}\vert}{r_c}, \qquad \frac{1}{r_{\sB}}=\frac{\vert\bN_{\sAB}\times\bn_{\sB}\vert}{r_c}
\ee
and 
\be \label{iden2}
\frac{r_c^2}{r_{\sA}r_{\sB}}=\bn_{\sA}.\bn_{\sB}-(\bN_{\! \sAB}.\bn_A)(\bN_{\! \sAB}.\bn_{\sB}),
\ee
it is easily seen that formulae (\ref{q1N})-(\ref{q3N}) are equivalent to the expressions of $q_1$, $q_2$, $q_3$ obtained in \cite{linet2}. 

Deriving now $c\partial {\cal T}^{(n)}/\partial r_{\sA}$ and $c\partial {\cal T}^{(n)}/\partial r_{\sB}$ for $n=1, 2, 3$ from (\ref{T1})-(\ref{T3}), and then substituting for the $q_n$ from  equations (\ref{q1N})-(\ref{q3N}) into (\ref{wlA3}) and (\ref{wlB3}), straightforward calculations lead to the explicit expressions of the light direction triples \index{light direction triple} up to order $G^3$. In fact, we may content ourselves with calculating $\widehat{\underline{\bm l}}_{\,e}^{(n)}$ or $ \widehat{\underline{\bm l}}_{\,r}^{(n)}$ since a relation as follows
\be \label{rec}
\widehat{\underline{\bm l}}_{\,r}^{(n)}(\bx_{\sA},\bx_{\sB})=-\widehat{\underline{\bm l}}_{\,e}^{(n)}(\bx_{\sB},\bx_{\sA})
\ee
results from (\ref{hlen})-(\ref{hlrn}) when the symmetry law ${\cal T}^{(n)}(\bx_{\sA} ,\bx_{\sB})={\cal T}^{(n)}(\bx_{\sB} ,\bx_{\sA})$ is taken into account. We get the proposition which follows.                           

\begin{proposition} \label{prophl123}
Under the assumption of proposition \ref{propT3}, the triples $\widehat{\underline{\bm l}}_{\,e}^{(n)}$ and $\widehat{\underline{\bm l}}_{\,r}^{(n)}$ are given for $n=1,2,3$ by
\begin{eqnarray}
&&\hspace{-2mm}\widehat{\underline{\bm l}}_{\,e}^{(1)}(\bx_{\sA},\bx_{\sB})=-\frac{(1+\gamma)m}{r_{\sA}}\bigg\lbrack \bN_{\sAB}+\frac{\vert\bn_{\sA}\times\bn_{\sB}\vert}{1+\bn_{\sA}.\bn_{\sB}}\bP_{\sAB}\bigg\rbrack, \label{exphle1} \\
&&\hspace{-2mm}\widehat{\underline{\bm l}}_{\,e}^{(2)}(\bx_{\sA},\bx_{\sB})=-\frac{m^2}{r_{\sA}^2}\bigg\lbrack\kappa -\frac{(1+\gamma)^2}{1+\bn_{\sA}.\bn_{\sB}}\bigg\rbrack\bN_{\sAB} \nonumber \\
&&\qquad\qquad\quad\;\;\, -\frac{m^2}{r_{\sA}^2}\frac{1}{\vert\bn_{\sA}\times\bn_{\sB}\vert}\bigg\lbrace\kappa\bigg\lbrack\frac{r_{\sA}}{r_{\sB}}-\bn_{\sA}.\bn_{\sB}\nonumber \\
&&\qquad\qquad\quad\;\;\,+\left(1-\frac{r_{\sA}}{r_{\sB}}\bn_{\sA}.\bn_{\sB}\right)\frac{\arccos\bn_{\sA}.\bn_{\sB}}{\vert\bn_{\sA}\times\bn_{\sB}\vert}\bigg\rbrack \nonumber \\
&&\qquad\qquad\quad\;\;\, -(1+\gamma)^2\left(1+\frac{r_{\sA}}{r_{\sB}}\right)\frac{1-\bn_{\sA}.\bn_{\sB}}{1+\bn_{\sA}.\bn_{\sB}}\bigg\rbrace\bP_{\sAB}, \label{exphle2} \\
&&\hspace{-2mm}\widehat{\underline{\bm l}}_{\,e}^{(3)}(\bx_{\sA},\bx_{\sB})=-\frac{m^3}{r_{\sA}^3}\bigg\lbrace\kappa_3 -\frac{(1+\gamma)\kappa}{1+\bn_{\sA}.\bn_{\sB}}\bigg\lbrack 1+\frac{r_{\sA}}{r_{\sB}}\nonumber \\
&&\qquad\qquad\quad\;\;\,+\left(1-\frac{r_{\sA}}{r_{\sB}}\bn_{\sA}.\bn_{\sB}\right)\frac{\arccos\bn_{\sA}.\bn_{\sB}}{\vert\bn_{\sA}\times\bn_{\sB}\vert} \bigg\rbrack \nonumber \\
&&\qquad\qquad\quad\;\;\, +\frac{(1+\gamma)^3}{(1+\bn_{\sA}.\bn_{\sB})^2}\bigg\lbrack 2+\frac{r_{\sA}}{r_{\sB}}(1-\bn_{\sA}.\bn_{\sB})\bigg\rbrack\bigg\rbrace\bN_{\sAB}\nonumber \\
&&\qquad\qquad\quad\;\;\, -\frac{m^3}{r_{\sA}^3}\frac{1}{\vert\bn_{\sA}\times\bn_{\sB}\vert}\bigg\lbrace\kappa_3(1-\bn_{\sA}.\bn_{\sB})\bigg\lbrack 1+\left(1+\frac{r_{\sA}}{r_{\sB}}\right)^2\frac{1}{1+\bn_{\sA}.\bn_{\sB}}\bigg\rbrack \nonumber \\
&&\qquad\qquad\quad\;\;\,-\frac{(1+\gamma)\kappa}{1+\bn_{\sA}.\bn_{\sB}}\bigg\lbrace\left(1+\frac{r_{\sA}}{r_{\sB}}\right)\left(\frac{r_{\sA}}{r_{\sB}}-\bn_{\sA}.\bn_{\sB}\right)\nonumber \\
&&\qquad\qquad\quad\;\;\, +\bigg\lbrack\left(1+\frac{r_{\sA}}{r_{\sB}}\right)^2(1-2\bn_{\sA}.\bn_{\sB}) \nonumber \\
&&\qquad\qquad\quad\;\;\,+\left(1+\frac{r_{\sA}}{r_{\sB}}\bn_{\sA}.\bn_{\sB}\right)(1+\bn_{\sA}.\bn_{\sB}) \bigg\rbrack\frac{\arccos\bn_{\sA}.\bn_{\sB}}{\vert\bn_{\sA}\times\bn_{\sB}\vert}\bigg\rbrace \nonumber \\
&&\qquad\qquad\quad\;\;\, +(1+\gamma)^3\frac{1-\bn_{\sA}.\bn_{\sB}}{1+\bn_{\sA}.\bn_{\sB}}\bigg\lbrack\left(1+\frac{r_{\sA}}{r_{\sB}}\right)^2\frac{2}{1+\bn_{\sA}.\bn_{\sB}}-\frac{r_{\sA}}{r_{\sB}} \bigg\rbrack \bigg\rbrace\bP_{\sAB} \nonumber \\
&&\label{exphle3}
\end{eqnarray}
and
\begin{eqnarray}
&&\hspace{-2mm}\widehat{\underline{\bm l}}_{\,r}^{(1)}(\bx_{\sA},\bx_{\sB})=-\frac{(1+\gamma)m}{r_{\sB}}\bigg\lbrack \bN_{\sAB}-\frac{\vert\bn_{\sA}\times\bn_{\sB}\vert}{1+\bn_{\sA}.\bn_{\sB}}\bP_{\sAB}\bigg\rbrack, \label{exphlr1} \\
&&\hspace{-2mm}\widehat{\underline{\bm l}}_{\,r}^{(2)}(\bx_{\sA},\bx_{\sB})=-\frac{m^2}{r_{\sB}^2}\bigg\lbrack\kappa -\frac{(1+\gamma)^2}{1+\bn_{\sA}.\bn_{\sB}}\bigg\rbrack\bN_{\sAB} \nonumber \\
&&\qquad\qquad\quad\;\;\, +\frac{m^2}{r_{\sB}^2}\frac{1}{\vert\bn_{\sA}\times\bn_{\sB}\vert}\bigg\lbrace\kappa\bigg\lbrack\frac{r_{\sB}}{r_{\sA}}-\bn_{\sA}.\bn_{\sB}\nonumber \\
&&\qquad\qquad\quad\;\;\,+\left(1-\frac{r_{\sB}}{r_{\sA}}\bn_{\sA}.\bn_{\sB}\right)\frac{\arccos\bn_{\sA}.\bn_{\sB}}{\vert\bn_{\sA}\times\bn_{\sB}\vert}\bigg\rbrack \nonumber \\
&&\qquad\qquad\quad\;\;\, -(1+\gamma)^2\left(1+\frac{r_{\sB}}{r_{\sA}}\right)\frac{1-\bn_{\sA}.\bn_{\sB}}{1+\bn_{\sA}.\bn_{\sB}}\bigg\rbrace\bP_{\sAB}, \label{exphlr2} \\
&&\hspace{-2mm}\widehat{\underline{\bm l}}_{\,r}^{(3)}(\bx_{\sA},\bx_{\sB})=-\frac{m^3}{r_{\sB}^3}\bigg\lbrace\kappa_3 -\frac{(1+\gamma)\kappa}{1+\bn_{\sA}.\bn_{\sB}}\bigg\lbrack 1+\frac{r_{\sB}}{r_{\sA}}\nonumber \\
&&\qquad\qquad\quad\;\;\,+\left(1-\frac{r_{\sB}}{r_{\sA}}\bn_{\sA}.\bn_{\sB}\right)\frac{\arccos\bn_{\sA}.\bn_{\sB}}{\vert\bn_{\sA}\times\bn_{\sB}\vert} \bigg\rbrack \nonumber \\
&&\qquad\qquad\quad\;\;\, +\frac{(1+\gamma)^3}{(1+\bn_{\sA}.\bn_{\sB})^2}\bigg\lbrack 2+\frac{r_{\sB}}{r_{\sA}}(1-\bn_{\sA}.\bn_{\sB})\bigg\rbrack\bigg\rbrace\bN_{\sAB}\nonumber \\
&&\qquad\qquad\quad\;\;\, +\frac{m^3}{r_{\sB}^3}\frac{1}{\vert\bn_{\sA}\times\bn_{\sB}\vert}\bigg\lbrace\kappa_3(1-\bn_{\sA}.\bn_{\sB})\bigg\lbrack 1+\left(1+\frac{r_{\sB}}{r_{\sA}}\right)^2\frac{1}{1+\bn_{\sA}.\bn_{\sB}}\bigg\rbrack \nonumber \\
&&\qquad\qquad\quad\;\;\, -\frac{(1+\gamma)\kappa}{1+\bn_{\sA}.\bn_{\sB}}\bigg\lbrace\left(1+\frac{r_{\sB}}{r_{\sA}}\right)\left(\frac{r_{\sB}}{r_{\sA}}-\bn_{\sA}.\bn_{\sB}\right)\nonumber \\
&&\qquad\qquad\quad\;\;\, +\bigg\lbrack\left(1+\frac{r_{\sB}}{r_{\sA}}\right)^2(1-2\bn_{\sA}.\bn_{\sB})\nonumber \\
&&\qquad\qquad\quad\;\;\, +\left(1+\frac{r_{\sB}}{r_{\sA}}\bn_{\sA}.\bn_{\sB}\right)(1+\bn_{\sA}.\bn_{\sB}) \bigg\rbrack\frac{\arccos\bn_{\sA}.\bn_{\sB}}{\vert\bn_{\sA}\times\bn_{\sB}\vert}\bigg\rbrace \nonumber \\
&&\qquad\qquad\quad\;\;\, +(1+\gamma)^3\frac{1-\bn_{\sA}.\bn_{\sB}}{1+\bn_{\sA}.\bn_{\sB}}\bigg\lbrack\left(1+\frac{r_{\sB}}{r_{\sA}}\right)^2\frac{2}{1+\bn_{\sA}.\bn_{\sB}}-\frac{r_{\sB}}{r_{\sA}} \bigg\rbrack \bigg\rbrace\bP_{\sAB}, \nonumber \\
&&\label{exphlr3}
\end{eqnarray}
respectively.
\end{proposition}

Using (\ref{iden1}), it is easily checked that equations (\ref{exphle1})-(\ref{exphle2}) and (\ref{exphlr1})-(\ref{exphlr2}) allow to recover expressions (38{\it a}) and (38{\it b}) obtained in \cite{teyssandier2} for the expansion of $\widehat{\underline{\bm l}}_{\,e}$ and $\widehat{\underline{\bm l}}_{\,r}$ up to the second order in $G$. Of course, (\ref{exphlr1}) and (\ref{exphlr2}) are equivalent to (\ref{l1r}) and (\ref{l2r}), respectively. On the other hand, formulae (\ref{exphle3}) and (\ref{exphlr3}) are new results.

\section{Light ray emitted at infinity} \label{rinfin}

A quasi-Minkowskian light ray \index{light ray} coming from infinity in an initial direction defined by a given unit vector $\bN_{e}$ and observed at a given point $\bx_{\sB}$ is a relevant limiting case for modelling a lot of astrometric measurements. According to a notation introduced in \cite{teyssandier2}, such a ray is denoted by $\Gamma_s(\bN_{e},\bx_{\sB})$. The corresponding null geodesic \index{null geodesic} is assumed to be a perturbation in powers of $G$ of the straight segment defined by the parametric equations
\be \label{param}
x^{0}_{(0)}(\lambda)=ct_{\sB}+\lambda r_c , \quad \bx_{(0)}(\lambda)=\lambda r_c\bN_e+\bx_{\sB}, \quad -\infty <\lambda\leq 0, 
\ee
where
\be \label{rcinf}
r_c=r_{\sB}\vert\bN_{e}\times\bn_{\sB}\vert .
\ee
Given the direction $\bN_{e}$, $\bx_{\sB}$ is supposed to satisfy the condition
\be \label{stli}
\vert \lambda r_c\bN_e+\bx_{\sB} \vert>r_h 
\ee
when $-\infty <\lambda\leq 0$ in order to ensure that the straight segment coming from infinity in the direction $\bN_{e}$ and ending at $\bx_{\sB}$ is entirely lying in ${\cal D}_h$. Condition (\ref{stli}) is the extension of condition (\ref{strl}) when $\bx_{\sA}$ is at infinity.  

To apply the results of section \ref{imp_par3}, let us consider a point $\bx_{\sA}$ lying on $\Gamma_s(\bN_{e},\bx_{\sB})$. It is clear that the part of $\Gamma_s(\bN_{e},\bx_{\sB})$ joining $\bx_{\sA}$ and $\bx_{\sB}$ coincides with a quasi-Minkowskian null geodesic \index{null geodesic} path $\Gamma_s(\bx_{\sA},\bx_{\sB})$. So, the impact parameters of $\Gamma_s(\bN_{e},\bx_{\sB})$ and $\Gamma_s(\bx_{\sA},\bx_{\sB})$ are equal. As a consequence, the coefficients $q_1, q_2$ and $ q_3$ can be obtained as functions of $\bN_{e}$ and $\bx_{\sB}$ by taking the limit of equations (\ref{q1N})-(\ref{q3N}) when $\bx_{\sA}$ recedes towards the source of the light ray \index{light ray} at infinity, i.e. when $r_{\sA}\rightarrow\infty$, $\bn_{\sA}\rightarrow -\bN_{e}$ and $\bN_{\sAB}\rightarrow\bN_{e}$. Using (\ref{rcinf}), and then taking into account that $\arccos \bn_{\sA}.\bn_{\sB}\rightarrow \pi-\arccos \bN_{e}.\bn_{\sB}$ when $\bn_{\sA}\rightarrow -\bN_{e}$, the following propositions can be stated.

\begin{proposition} \label{propqinf}
Let $\bN_{e}$ be a unit vector and $\bx_{\sB}$ a point in ${\cal D}_h$ fulfilling condition (\ref{stli}). The impact parameter of a quasi-Minkowskian light ray \index{impact parameter of a light ray} emitted at infinity in the direction $\bN_{e}$ and arriving at $\bx_{\sB}$ is given by expansion (\ref{expb}), where $r_c$ is expressed by (\ref{rcinf}) and the coefficients $q_1$, $q_2$ and $q_3$ are yielded by\footnote{Note that equation (114) yielding $q_2$ in \cite{linet2} contains 
an extra factor $\vert\bN_{e}\times\bn_{\sB}\vert$ due to a typographic mistake. See the corrigendum quoted in \cite{linet2}.}
\begin{eqnarray} 
& & q_1(\bN_{e},\bx_{\sB})= ( 1+\gamma ) \frac{\vert\bN_{e}\times\bn_{\sB}\vert}{1-\bN_{e}.\bn_{\sB}},  \label{q1inf} \\
& & q_2(\bN_{e},\bx_{\sB})=\kappa \left[ 1+\bN_{e}.\bn_{\sB}\frac{\pi-\arccos \bN_{e}.\bn_{\sB}}{\vert\bN_{e}\times\bn_{\sB}\vert}\right]-(1+\gamma )^2\frac{1+\bN_{e}.\bn_{\sB}}{1-\bN_{e}.\bn_{\sB}},\nonumber \\
&& \label{q2inf}\\
& & q_3 (\bN_{e},\bx_{\sB})=\frac{\vert\bN_{e}\times\bn_{\sB}\vert}{1-\bN_{e}.\bn_{\sB}} \bigg\lbrace \kappa_3\left( 1+\bN_{e}.\bn_{\sB}\right)+2(1+\gamma )^3\frac{1+\bN_{e}.\bn_{\sB}}{1-\bN_{e}.\bn_{\sB}} \nonumber \\
& &\qquad\qquad\quad\;\;\;\, -(1+\gamma)\kappa\left[1+\left(1+2\bN_{e}.\bn_{\sB} \right)  
\frac{\pi-\arccos \bN_{e}.\bn_{\sB}}{\vert\bN_{e}\times\bn_{\sB}\vert} \right] \bigg\rbrace . \label{q3inf}
\end{eqnarray} 
\end{proposition}

\begin{proposition} \label{prophlinf}
For the ray \index{light ray} considered in proposition \ref{propqinf}, the light direction triple \index{light direction triple} at point $\bx_{\sB}$ is determined up to the third order by 
\begin{eqnarray}
&&\widehat{\underline{\bm l}}_{\,r}^{(1)}(\bN_{e},\bx_{\sB})=-\frac{(1+\gamma)m}{r_{\sB}}\bigg\lbrack \bN_{e}-\frac{\vert\bN_{e}\times\bn_{\sB}\vert}{1-\bN_{e}.\bn_{\sB}}\bP_{e}\bigg\rbrack, \label{hlrinf1} \\
&&\widehat{\underline{\bm l}}_{\,r}^{(2)}(\bN_{e},\bx_{\sB})=-\frac{m^2}{r_{\sB}^2}\bigg\lbrack\kappa -\frac{(1+\gamma)^2}{1-\bN_{e}.\bn_{\sB}}\bigg\rbrack\bN_{e} \nonumber \\
&&\qquad\qquad\qquad\;\; +\frac{m^2}{r_{\sB}^2}\frac{1}{\vert\bN_{e}\times\bn_{\sB}\vert}\bigg\lbrace\kappa\bigg\lbrack\bN_{e}.\bn_{\sB}+\frac{\pi -\arccos\bN_{e}.\bn_{\sB}}{\vert\bN_{e}\times\bn_{\sB}\vert}\bigg\rbrack \nonumber \\
&&\qquad\qquad\qquad\;\; -(1+\gamma)^2\frac{1+\bN_{e}.\bn_{\sB}}{1-\bN_{e}.\bn_{\sB}}\bigg\rbrace\bP_{e}, \label{hlrinf2} \\
&&\widehat{\underline{\bm l}}_{\,r}^{(3)}(\bN_{e},\bx_{\sB})=-\frac{m^3}{r_{\sB}^3}\bigg\lbrace\kappa_3 -\frac{(1+\gamma)\kappa}{1-\bN_{e}.\bn_{\sB}}\bigg\lbrack 1+\frac{\pi -\arccos\bN_{e}.\bn_{\sB}}{\vert\bN_{e}\times\bn_{\sB}\vert} \bigg\rbrack \nonumber \\
&&\qquad\qquad\qquad\;\; +\frac{2(1+\gamma)^3}{(1-\bN_{e}.\bn_{\sB})^2} \bigg\rbrace\bN_{e}\nonumber \\
&&\qquad\qquad\qquad\;\; +\frac{m^3}{r_{\sB}^3}\frac{1}{\vert\bN_{e}\times\bn_{\sB}\vert}\bigg\lbrace\kappa_3 \frac{1+\bN_{e}.\bn_{\sB}}{1-\bN_{e}.\bn_{\sB}} (2-\bN_{e}.\bn_{\sB})\nonumber \\
&&\qquad\qquad\qquad\;\; -\frac{(1+\gamma)\kappa}{1-\bN_{e}.\bn_{\sB}}\bigg\lbrack\bN_{e}.\bn_{\sB}+(2+\bN_{e}.\bn_{\sB})\frac{\pi -\arccos\bN_{e}.\bn_{\sB}}{\vert\bN_{e}\times\bn_{\sB}\vert}\bigg\rbrack \nonumber \\
&&\qquad\qquad\qquad\;\; +2(1+\gamma)^3\frac{1+\bN_{e}.\bn_{\sB}}{(1-\bN_{e}.\bn_{\sB})^2} \bigg\rbrace\bP_{e}, \label{hlrinf3}
\end{eqnarray}
where $\bP_{e}$ is defined as
\be \label{Pe}
\bP_{e}=\left(\frac{\bN_{e}\times\bn_{\sB}}{\vert\bN_{e}\times\bn_{\sB}\vert}\right)\times\bN_{e}.
\ee
\end{proposition}

Exactly as in the case where the emission point is located at a finite distance from the origin, formula (\ref{hlrinf3}) is  new, whereas (\ref{hlrinf1}) and (\ref{hlrinf2}) are equivalent to the expressions of the direction triples up to the second order derived in \cite{teyssandier2}.

\section{Enhanced terms in ${\cal T}^{(1)}$, ${\cal T}^{(2)}$ and ${\cal T}^{(3)}$} \label{Suit3rd}

In the present work, the time transfer function \index{time transfer function} ${\cal T}$ is obtained in the form of an asymptotic expansion in power series in $G$ (or $m$) provided that condition (\ref{strl}) is met. However, it is clear that the physical reliability of this expansion requires that inequalities as follow 
\be \label{CTn}
\left\vert{\cal T}^{(n)}( \bx_{\sA}, \bx_{\sB})\right\vert \ll \left\vert{\cal T}^{(n-1)}( \bx_{\sA}, \bx_{\sB})\right\vert 
\ee
are satisfied for any $n\geq 1$, with ${\cal T}^{(0)}( \bx_{\sA}, \bx_{\sB})$ being conventionally defined as  
\[
{\cal T}^{(0)}( \bx_{\sA}, \bx_{\sB})=\frac{1}{c}\vert \bx_{\sB}-\bx_{\sA}\vert.
\]

The results obtained in the previous section enable us to find the conditions ensuring inequalities (\ref{CTn}) for $n=1, 2, 3$. It is clear that the magnitude of the functions given by (\ref{T1})-(\ref{T3}) may be extremely large when points $\bx_{\sA}$ and $\bx_{\sB}$ are located in almost opposite directions. This behaviour corresponds to the `enhanced terms' determined up to $G^2$ for the deflection of light \index{deflection of light} in \cite{klioner1} and up to $G^3$ for the time transfer function in \cite{ashby}. Indeed, it is straightforwardly derived from (\ref{rc}) that
\be \label{asyn}
\frac{1}{1+\bn_{\sA}.\bn_{\sB}}\sim \frac{2 r_{\sA}^2 r_{\sB}^2}{(r_{\sA} +r_{\sB})^2}\frac{1}{r_c^2}
\ee
when $1+\bn_{\sA}.\bn_{\sB}\rightarrow 0$. Using this relation to eliminate $1+\bn_{\sA}.\bn_{\sB}$, the following proposition is easily deduced from (\ref{T1})-(\ref{T3}). 

\begin{proposition} \label{propAsymp}
When $\bx_{\sA}$ and $\bx_{\sB}$ tend to be located in opposite directions ({\it i.e.} $1+\bn_{\sA}.\bn_{\sB}\rightarrow 0$), the first three perturbation terms in the time transfer function \index{time transfer function} are enhanced according to the asymptotic expressions
\begin{eqnarray}
&&{\cal T}^{(1)}_{enh}(\bx_{\sA}, \bx_{\sB})\sim \frac{(1+\gamma)m}{c}\ln\left(\frac{4 r_{\sA} r_{\sB}}{r^2_c}\right), \label{asT1}\\
&&{\cal T}^{(2)}_{enh}(\bx_{\sA}, \bx_{\sB})\sim -2 \frac{(1+\gamma)^2 m^2}{c(r_{\sA}+r_{\sB})} \frac{r_{\sA} r_{\sB}}{r^2_c}, \label{asT2}\\
&&{\cal T}^{(3)}_{enh}(\bx_{\sA}, \bx_{\sB})\sim 4\frac{(1+\gamma)^3 m^3}{c(r_{\sA}+r_{\sB})^2} \left(\frac{r_{\sA} r_{\sB}}{r^2_c}\right)^2. \label{asT3}
\end{eqnarray}
\end{proposition}

These expressions confirm the formulae obtained in \cite{ashby} by a different method. It is worthy noticing that, at least up to $G^3$, $\gamma$ is the only post-Newtonian parameter involved in the enhanced terms. When $\bx_{\sA}$ and $\bx_{\sB}$ tend to be located in opposite directions, the asymptotic behavior of each function ${\cal T}^{(n)}_{enh}$ is such that 
\be \label{CasTn}
\left\vert{\cal T}^{(n)}_{enh}(\bx_{\sA}, \bx_{\sB})\right\vert \lesssim k_n \frac{2(1+\gamma) m}{r_{\sA}+r_{\sB}} \frac{r_{\sA} r_{\sB}}{r^2_c} \left\vert{\cal T}^{(n-1)}_{enh}(\bx_{\sA}, \bx_{\sB})\right\vert
\ee
for $n=1,2,3$, with $k_1=2$, $k_2=k_3=1$ and ${\cal T}^{(0)}_{enh}(\bx_{\sA}, \bx_{\sB})\sim r_{\sA}+r_{\sB}$. For $n=3$, the formula (\ref{CasTn}) is straightforwardly derived from (\ref{asT2}) and (\ref{asT3}) (the symbol $\lesssim$ could be replaced by $\sim$). For $n=1$, the formula results from the fact that $\ln x<x$  for any $x>0$. Lastly, for $n=2$, (\ref{CasTn}) obviously follows from the fact that $\ln (4 r_{\sA} r_{\sB}/r^2_c) \rightarrow \infty$ when $1+\bn_{\sA}.\bn_{\sB}\rightarrow 0$.

It results from (\ref{CasTn}) that inequalities (\ref{CTn}) are satisfied for $n=1,2,3$ as long as the zeroth-order distance of closest approach is such that a condition as follows
\be \label{asyp}
\frac{2 m}{r_{\sA}+r_{\sB}} \frac{r_{\sA} r_{\sB}}{r^2_c} \ll 1
\ee
is fulfilled. This inequality coincides with the condition ensuring the validity of the asymptotic expansions obtained in \cite{ashby}. It may be expected that (\ref{asyp}) is sufficient to ensure inequality (\ref{CTn}) at any order.

It results from the definition of $r_c$ that condition (\ref{asyp}) is equivalent to 
\[
r_c \gg 2m\frac{\vert\bx_{\sB}-\bx_{\sA}\vert}{r_{\sA}+r_{\sB}}\frac{1}{\vert\bn_{\sA}\times\bn_{\sB}\vert}.
\]
When $1+\bn_{\sA}.\bn_{\sB}\rightarrow 0$, this inequality implies 
\be \label{rcg}
r_c \gg \frac{m}{\sqrt{1+\bn_{\sA}.\bn_{\sB}}},
\ee
which in turn implies $r_c \gg m$. This last inequality means that condition (\ref{strl}) is met when inequality (\ref{asyp}) is satisfied\footnote{Note that the reciprocal is not true.}. The full expressions of ${\cal T}^{(1)}$, ${\cal T}^{(2)}$ and ${\cal T}^{(3)}$ obtained by the procedures developed in the present work can therefore be considered as reliable in a close superior conjunction as long as inequalities (\ref{CTn}) hold.

To finish, it is worth noticing that condition (\ref{asyp}) applied to a close superior conjunction is equivalent to an inequality as follows
\be \label{vasy}
\pi - \arccos \bn_{\sA}.\bn_{\sB} \gg \sqrt{\frac{2m(r_{\sA}+r_{\sB})}{r_{\sA} r_{\sB}}}
\ee
when (\ref{rc}) is taken into account. This last inequality clearly indicates that our procedures cannot be straightforwardly applied to the gravitational lensing configurations.

\section{Application to some \index{solar system experiments} solar system experiments} \label{Apexp}

Condition (\ref{asyp}) is fulfilled in experiments performed with photons exchanged between a spacecraft in the outer solar system and a ground station. Indeed, noting that 
\[
\frac{m}{r_c} \frac{r_{\sB}}{r_c}<\frac{2 m}{r_{\sA}+r_{\sB}} \frac{r_{\sA} r_{\sB}}{r^2_c}<2\frac{m}{r_c} \frac{r_{\sB}}{r_c}
\]
holds if $r_{\sA}>r_{\sB}$, replacing $m$ by half the Schwarzschild radius of the Sun, $m_{\odot}$, and then putting $r_{\sB}=$ 1 au, we find that inequalities
\be \label{exasyp}
4.56 \times 10^{-4} \times \frac{R_{\odot}^2}{r_c^2}<\frac{2 m_{\odot}}{r_{\sA}+r_{\sB}} \frac{r_{\sA} r_{\sB}}{r^2_c}<
9.12\times 10^{-4} \times \frac{R_{\odot}^2}{r_c^2}
\ee
hold if $r_{\sA}>r_{\sB}$, with $R_{\odot}$ denoting the radius of the Sun. We put $R_{\odot}=6.96\times 10^8$ m. The other numerical parameters of the Sun used throughout this section are taken from \cite{iers}. 

\begin{table}
\begin{center}
\begin{tabular}{l l l l l l}   \hline
\vspace{0.1 mm}$r_c/R_{\odot}$ & \vspace{0.1 mm}$\; \; | {\cal T}_{S}^{(1)} |$ & \vspace{0.1 mm}$\; \; \; \; {\cal T}_{J_2}^{(1)}$ & 
\vspace{0.1 mm}$\; \; \; \; \; {\cal T}_{enh}^{(2)}$ & \vspace{0.1 mm}$\; \; \; \; {\cal T}_{\kappa}^{(2)}$ & \vspace{0.1 mm}$\; \; \; \; {\cal T}_{enh}^{(3)}$ \\ \hline
$\; \; \; \; 1$ & $\; \; \; \; 10$ & $\; \; \; \; \; 2$ & $-17616$ & $\; \; 123$ & $\; \; \; \; 31.5$ \\
$\; \; \; \; 2$ & $\; \; \; \; \; \; 5$ & $\; \; \; \; \; 0.5$ & $\; \; -4404$ & $\; \; \; \; 61.5$ & $\; \; \; \; \; \; 2$ \\ 
$\; \; \; \; 5$ & $\; \; \; \; \; \; 2$ & $\; \; \; \; \; 0.08$ & $\; \; \; \; -704.6$ & $\; \; \; \; 24.6$ & $\; \; \; \; \; \; 0.05$ \\ \hline
\end{tabular}
\caption{\small Numerical values in ps of the main stationary contributions to the light travel time \index{light travel time} in the solar system for various values of $r_c/R_{\odot}$. In each case, $r_{\sA}=50$ au and $r_{\sB}=1$ au. The parameters $\gamma$ and $\kappa$ are taken as $\gamma=1$ and $\kappa=15/4$, respectively. For the numerical estimates of $\vert{\cal T}^{(1)}_{S}\vert$ and ${\cal T}^{(1)}_{J_2}$, the light ray is assumed to propagate in the equatorial plane of the Sun. The dynamical effects due to the planetary perturbations are not taken into account.\label{table1}}
\end{center}
\end{table} 

Formulae (\ref{asT1})-(\ref{asT3}) enable us to discuss the relevance of the terms ${\cal T}^{(2)}_{enh}$ and ${\cal T}^{(3)}_{enh}$ in a proposed mission like SAGAS, for instance. Indeed, this project plans to measure the parameter $\gamma$ up to an accuracy reaching $10^{-8}$ with light rays travelling between a spacecraft moving in the outer solar system and the Earth. For $r_{\sA}=50$ au and $r_{\sB}=1$ au, the travel time of a ray passing in close proximity to the Sun (conjunction) is about $2.54\times 10^4$ s. It follows from (\ref{asT1}) that ${\cal T}^{(1)}$ is decreasing from 158 $\mu$s to 126 $\mu$s when $r_c$ varies from $R_{\odot}$ to $5 R_{\odot}$. As a consequence, reaching an accuracy of $10^{-8}$ on the measurement of $\gamma$ requires to determine the light travel time \index{light travel time} with an accuracy of 0.7 ps. The numerical values of the respective contributions of ${\cal T}^{(2)}_{enh}$ and ${\cal T}^{(3)}_{enh}$ are indicated in table \ref{table1}. It is clear that the contribution of the enhanced term of order $G^3$ is larger than 2 ps when $r_c < 2R_{\odot}$. The same order of magnitude for ${\cal T}^{(3)}_{enh}$ may be expected in other proposed missions like ODYSSEY, LATOR or ASTROD. 

The above discussion also reveals that an experiment like SAGAS would enable to determine the post-post-Newtonian parameter $\kappa$ with a relative precision amounting to $7 \times 10^{-3}$. In the solar system, indeed, the term proportional to $\kappa$ in (\ref{T2}) yields the asymptotic contribution
\be \label{asT2k}
{\cal T}^{(2)}_{\kappa}(\bx_{\sA}, \bx_{\sB}) \sim \frac{\kappa \pi m_{\odot}^2}{c r_c}
\ee
when (\ref{asyn}) holds. For a ray grazing the Sun ($r_c=R_{\odot}$), one has ${\cal T}^{(2)}_{\kappa}\approx 123$ ps if $\kappa =15/4$. Hence the conclusion.

Before closing this study, it is worthy of note that the first-order contribution ${\cal T}^{(1)}_{S}$ to the time transfer function \index{time transfer function} due to the gravitomagnetic effect of the solar rotation may be compared with the third-order enhanced term. Indeed, it is easily inferred from equation (62) in \cite{linet} that for a ray travelling in the equatorial plane of the Sun
\be \label{asT1S}
\left\vert{\cal T}^{(1)}_{S}(\bx_{\sA}, \bx_{\sB})\right\vert \sim \frac{2(1+\gamma)G S_{\odot}}{c^4 r_c}
\ee
when (\ref{asyn}) is checked, with $S_{\odot}$ being the angular momentum of the Sun. According to helioseismology, we can take $S_{\odot} \approx 2\times 10^{41}$ kg m$^2$ s$^{-1}$ (see, e.g., \cite{komm}). So, in the case where $r_c=R_{\odot}$, we have $\vert{\cal T}^{(1)}_{S}(\bx_{\sA}, \bx_{\sB})\vert\approx 10$ ps. Furthermore, the contribution ${\cal T}^{(1)}_{J_2}$ due to the solar quadrupole moment $J_{2\odot}$ must also be considered for rays grazing the Sun. Using equation (24) in \cite{leponcin2} for a ray travelling in the equatorial plane gives
\be \label{asJ2}
{\cal T}^{(1)}_{J_2}(\bx_{\sA}, \bx_{\sB}) \sim \frac{(1+\gamma)m_{\odot}}{c} J_{2\odot} \frac{R_{\odot}^2}{r_c^2}.
\ee
Taking $J_{2\odot} \approx2\times 10^{-7}$ and putting $r_c=R_{\odot}$, (\ref{asJ2}) leads to ${\cal T}^{(1)}_{J_2}\approx2$ ps.

\section{Concluding remarks} \label{concl}

Two methodologies enabling us to determine the time transfer function \index{time transfer function} up to any given order of approximation in a static, spherically symmetric spacetime \index{spherically symmetric spacetime} have been presented. The corresponding procedures are natively adapted to the case where both the emitter and the receiver of light rays are located at a finite distance from the origin of the spatial coordinates. These procedures lead to identical expressions for the time transfer function up to the order $G^3$ (see eqs. (\ref{T1})-(\ref{T3})). This coincidence is a new result. The reliability of the expression obtained for ${\cal T}^{(3)}$ in \cite{linet2} is thus confirmed.

The procedure set out in section \ref{Determ2} presents the advantage of exclusively involving elementary integrations which can be performed with any symbolic computer program, whatever the order of approximation. It is very likely that the procedure applied in section \ref{Determ1} has the same property, even if it is not so easy to prove.

It must be emphasized that the explicit determination of the time transfer function up to the third order does not reduce to a purely mathematical improvement. This determination brings a rigorous proof of the existence of a third-order enhanced contribution to the time transfer function for light rays grazing the Sun. The enhanced term in ${\cal T}^{(3)}$ must be taken into account for determining the post-Newtonian parameter $\gamma$ at a level of accuracy of $10^{-8}$ in \index{solar system experiments} solar system experiments. It is worth noticing that for light rays almost grazing the Sun, this enhanced term is larger than the first-order Lense-Thirring effect due to the spinning of the Sun and than the first-order contribution due to the solar mass-quadrupole.

The light direction triples \index{light direction triple} $\widehat{\underline{\bl}}_{\,e}$ and $\widehat{\underline{\bl}}_{\,r}$ are now fully calculated up to the third order in $G$ (see eqs. (\ref{exphle1})-(\ref{exphlr3}) for the generic case and (\ref{hlrinf1})-(\ref{hlrinf3}) for a light ray emitted at infinity).  As a consequence, the frequency shift \index{frequency shift} between two observers could be determined up to the order $G^3$ by means of formula (\ref{nuABl}).

To finish, it may be noted that the calculations of the time transfer function performed here with the second procedure could be easily extended to quasi-Minkowskian light rays propagating in the equatorial plane of an axisymmetric, rotating body having a nonzero mass-quadrupole. This methodology could be of interest for studying the light propagation in a Kerr metric, for example.


\section*{Acknowledgements}

We are grateful to Bernard Linet for very hepful discussions and indebted to Olivier Minazzoli for his precious remarks.



\end{document}